\documentclass[aps,prb,showpacs,showkeys,superscriptaddress,10pt,reprint,preprintnumbers]{revtex4-1}

\usepackage[utf8]{inputenc}	
\usepackage[T1]{fontenc}
\usepackage[english]{babel}

\usepackage{amsmath}
\usepackage{amsfonts}
\usepackage{latexsym}
\usepackage{amssymb} 
\usepackage{MnSymbol} 
\usepackage{bbm} 
\usepackage{multirow}
\usepackage{textcomp}

\usepackage[pdftex]{graphicx} 
\graphicspath{{./pictures/}}

\newcommand{\bref}[1]{(\ref{#1})}

\newcommand{\bk}{{\bf k}}

\newcommand{\bS}{{\bf \hat S}}
\newcommand{\itext}[1]{\text{\it #1}}
\newcommand{\tJU}{$t$--$J$--$U$}

\usepackage[usenames,dvipsnames]{xcolor}

\begin{document}


\title{$d$-wave superconductivity and its coexistence with antiferromagnetism in $t$--$J$--$U$ model:
Statistically consistent Gutzwiller approach}

\author{M. Abram}
\email{marcin.abram@uj.edu.pl}

\author{J. Kaczmarczyk}
\affiliation{Marian Smoluchowski Institute of Physics, Jagiellonian University, Reymonta 4, 30-059 Kraków, Poland}

\author{J. J\k{e}drak}
\affiliation{Faculty of Non-Ferrous Metals, AGH University of Science and Technology, Al. Mickiewicza 30, 30-059 Kraków, Poland}

\author{J. Spa\l{}ek}
\email{ufspalek@if.uj.edu.pl}
\affiliation{Marian Smoluchowski Institute of Physics, Jagiellonian University, Reymonta 4, 30-059 Kraków, Poland}
\affiliation{Academic Center for Materials and Nanotechnology, Faculty of Physics and Applied Computer Science, AGH University of Science and Technology, Reymonta 19, 30-059 Krak\'ow, Poland}

\date{\today}

\preprint{It's ArXiv coppy. see PRB {\bf 88}, 094502 instead, DOI:10.1103/PhysRevB.88.094502}

\begin{abstract}

We discuss the coexistence of antiferromagnetism and $d$-wave superconductivity within the so-called statistically-consistent Gutzwiller approximation (SGA) applied to the $t$--$J$--$U$ model. In this approach, the averages calculated in a self-consistent manner coincide with those determined variationally. Such consistency is not guaranteed within the standard renormalized mean field theory. With the help of SGA, we show that for the typical value $J/|t| = 1/3$, coexistence of antiferromagnetism (AF) and superconductivity (SC) appears only for $U/|t| > 10.6$ and in a very narrow range of doping $(\delta \lesssim 0.006)$ in the vicinity of the Mott insulating state, in contrast to some previous reports.
In the coexistent AF+SC phase, a staggered spin-triplet component of the superconducting gap appears also naturally; its value is very small.
\noindent {\bf }
\end{abstract}



\maketitle

\section{Introduction: rationale for \tJU\ model}

High-temperature superconductivity in cuprates is often described within the effective $t$--$J$ model\cite{Spatek1977-PhysicaBC.86-88.375, *ChaoSpalekOles1977-JPhysC.10.L271, Spalek2007-ActaPhysPolonA.111.409}
(for a preliminary treatment of the topic c.f. also Ref.~\onlinecite{Anderson1988}).
The model justifies a number of experimental results,
such as superconductivity's dome-like shape on doping-temperature phase diagram,\cite{Dagotto1994-RevModPhys.66.763}
non-Fermi-liquid behavior of the normal state for underdoped and optimally doped systems,\cite{Sarker1994-PhysC, *Sarker2010-PRB.82.014504, Johnson2001-JournOfElectrSpectros.117.153, Lee2006-RevModPhys.78.17}
the disappearance of the pairing gap magnitude in the antiferromagnetic state (albeit only at the doping $\delta=0$),\cite{Lee2006-RevModPhys.78.17, Imada1998-RevModPhys.70.1039}
and the doping dependence of the photoemission spectrum in the antinodal direction.\cite{Damascelli2003-RevModPhys.75.473, Tanaka2006-Science.314.1910}
All of these features represent an attractive starting point for further analysis (cf.~Ref.~\onlinecite{Scalapino2012-RevModPhys.84.1383}).

In the effective $t$--$J$ model, the value of the kinetic exchange integral $J_{ij}$ does not necessarily coincide with the value $J_{ij} = 4 t_{ij}^2 / U$ obtained perturbationally from the Hubbard model.\cite{Spalek2007-ActaPhysPolonA.111.409} Instead, it expresses an effective coupling between the copper spins in mixed copper-oxygen $3d$--$2p$ holes.\cite{Zhang1988-PhysRevB.37.3759} Therefore, one may say that the values of the hopping integral $t_{ij}$ and that of antiferromagnetic exchange $J_{ij}$ in that model are practically independent. 
Typically, the ratio $|t|/J \approx 3$ is taken and corresponds to the value $U/(8 |t|) = 1.5$ in the context of the two-dimensional Hubbard model. However, after introducing the bare bandwidth $W=8|t|$ in the tight-binding approximation for a square lattice, we obtain the ratio $U/W = 1.5$, which is not sufficiently large for the transformation of the original Hubbard model into the $t$--$J$ model to be valid in the low order. In that situation, we are, strictly speaking not within the strong correlation limit $U/W \gg 1$, in which the $t$--$J$ model was originally derived.\cite{Spatek1977-PhysicaBC.86-88.375, *ChaoSpalekOles1977-JPhysC.10.L271, Spalek2007-ActaPhysPolonA.111.409}

In order to account properly for the strong electronic correlations (the bare Hubbard parameter $U$ for Cu$^{2+}$ ion is $8\mbox{-}10\ \mbox{eV} \gg W \approx 2-3\ \mbox{eV}$), we can add the Hubbard term $U \sum_i \hat n_{i\uparrow} \hat n_{i\downarrow}$ to the $t$--$J$ model. In this manner, we consider the exchange integral $J_{ij}$ in this still-effective single-band model as coming from the full superexchange involving the oxygen ions rather than from the effective kinetic exchange only (for critical overview c.f. Ref.~\onlinecite{Zaanen19990-JSSC.88.8}). This argument may be regarded as one of the justifications for introducing the $t$--$J$--$U$ model, first used by Daul,\cite{Daul2000-PRL.84.4188} Basu,\cite{Basu2001-PRB.63.100506} and Zhang\cite{Zhang2003-prl} (cf. Ref.~\onlinecite{Xiang2009-PRB.79.014524}, where comprehensive justification of the $t$--$J$--$U$ is provided).

There is an additional reason for the $t$--$J$--$U$ model applicability to the cuprates.
Namely, in the starting, bare configuration of CuO$_2^{2-}$ structural unit, the hybridization between the antibonding  $2p_{\sigma}$ states due to oxygen and one-hole $(3d^{9})$ states due to Cu is strong, with the hybridization matrix element $|V_{\langle im \rangle}| \sim 1.5$ eV.
Therefore, the hybridization contribution to the hole state itinerancy, at least on the single-particle level, is essential and hence the effective $d$--$d$ (Hubbard) interaction is substantially reduced.
In effect, we may safely assume that $U \gtrsim W$ instead $U \gg W$.
In this manner, the basic simplicity of the single-band model is preserved, as it provides not only the description of the strongly correlated metallic state close to the Mott insulating limit, but also reduces to the correct limit of the Heisenberg magnet of spin $1/2$ with strong antiferromagnetic exchange integral $J \approx 0.1$ eV in the absence of holes (the Mott-Hubbard insulating state). Last but not least, within the present model we can study the limit $U \rightarrow \infty$ and compare explicitly the results with those of canonical $t$--$J$ model.

Antiferromagnetism (AF) and superconductivity (SC) can coexist in the
{\it electron doped} cuprates,\cite{Yu2007-PhysRevB.76.020503, Armitage2010-RevModPhys.82.2421}
but in the {\it hole-doped} cuprates the two phases are usually separated (cf. e.g. the review of Dagotto\cite{Dagotto1994-RevModPhys.66.763}). However, in the late 1990s, reports of a possible coexistence in the cuprates appeared, first vague (cf. Ref.~\onlinecite{Kimura1999-PhysRevB.59.6517} (La$_{2-x}$Sr$_x$Cu$_{1-y}$Zn$_y$O$_4$)), then more convincing (cf. e.g. Ref.~\onlinecite{Lee1999-PhysRevB.60.3643} (La$_2$CuO$_{4+y}$), Ref.~\onlinecite{Sidis2001-PhysRevLett.86.4100} (Y\!Ba$_2$Cu$_3$O$_{6.5}$) or Ref.~\onlinecite{Hodges2002-PhysRevB.66.020501} (YBa$_2$(Cu$_{0.987}$Co$_{0.013}$)$_3$O$_{y+\delta}$)).
Other systems, where the coexistence has been reported, are organic superconductors,\cite{Kawamoto2008-PhysRevB.77.224506} heavy-fermions systems,\cite{Mathur1998-2010-Nature.394.39} iron-based superconductors such as Ba(Fe$_{1-x}$Ru$_x$)$_2$As$_2$ (Ref.~\onlinecite{Ma2012-PhysRevLett.109.197002}), Ba$_{0.77}$K$_{0.23}$Fe$_2$As$_2$ (Ref.~\onlinecite{Li2012-PhysRevB.86.180501}), Ba(Fe$_{1-x}$Co$_x$)$_2$As$_2$ (Ref.~\onlinecite{Marsik2010-PhysRevLett.105.057001, Bernhard2012-PhysRevB.86.184509}), as well as graphene bilayer systems (cf. Ref.~\onlinecite{Milovanovic2012-PhysRevB.86.195113}).

Our purpose is to undertake a detailed analysis of the paired (SC) state within the $t$--$J$--$U$ model and its coexistence with the two-sublattice antiferromagnetism in two dimensions.
Detailed studies of the \tJU\ model have been carried out by Zhang,\cite{Zhang2003-prl} Gan,\cite{Gan2005-prb, Gan2005-prl} and Bernevig\cite{Bernevig-PhysRevLett.91.147003} who described a transition from gossamer\cite{Laughlin2006} and $d$-wave\cite{Harlingen1995-RevModPhys.67.515, Tsuei2000-RevModPhys.72.969} superconductivity to the Mott insulator.
However, the existence of AF order was not considered in those studies. Some attempts to include AF order were made by Yuan\cite{Yuan2005-prb} and Heiselberg,\cite{Heiselberg2009-pra} and very recently by Voo\cite{Voo-2011-JPhysCondensMatter.23.495602} and Liu,\cite{Fen2012-CommTheorPhys} but in all those works one can question the authors' approach. Specifically, the equations used do not guarantee self-consistency, i.e. the mean-field averages introduced in a self-consistent manner do not match those determined variationally.\cite{Jedrak2010-arXiv}
We show that the above problem that  appears in the
Renormalized Mean Field Theory (RMFT) formulation, can be overcome by introducing constraints that ensure the statistical consistency between the two above ways of determining mean-field values. 
This is the principal concept of our {\it statistically consistent Gutzwiller approach} (SGA).\cite{Jedrak2010-prb, Jedrak2011-PhD}

Using SGA we obtain that AF phase is stable only in the presence of SC in a very narrow region close to the Mott-Hubbard insulating state, corresponding to the half-filled (undoped) situation.
Additionally, in this AF-SC coexisting phase, a small staggered spin-triplet component of the superconducting gap appears naturally, in addition to the predominant spin-singlet component.

The structure of the paper is as follows. In Sec.~II, we define the model and provide definitions of the mean-field parameters. In Sec.~III, we introduce the constraints with the corresponding Lagrange multipliers to guarantee the consistency of the self-consistent and the variational procedures of determining the mean-field parameters. 
The full minimization procedure is also outlined there. In Sec.~IV, we discuss the numerical results, as well as provide the values of the introduced Lagrange multipliers. 
In Sec.~V, we summarize our results and compare them with those of other studies.
In Appendix A, we discuss the general form of the hopping amplitude and the superconducting gap, as well as some details of the analytic calculations required to determine the ground state energy.
In Appendixes B and C, we show some details of our calculations.
In Appendix D, we present an alternative and equivalent procedure of introducing the Lagrange multipliers to that presented in the main text.
In Appendix E, we list representative values of the parameters calculated for different phases.

\section{\tJU\ Model and effective single-particle Hamiltonian}

We start from the \tJU\ model as represented by the Hamiltonian\cite{Zhang2003-prl, Gan2005-prb, Gan2005-prl}
\begin{equation}
   \mathcal{\mathcal{\hat H}} = \ t \sum_{\left\langle i\, j \right\rangle,\, \sigma} \left( \hat c^\dagger_{i\sigma} \hat c_{j \sigma} + \text{H.c.} \right) + J \sum_{\left\langle i\, j \right\rangle} \bS_i \cdot \bS_{j} + U \sum_i \hat n_{i\uparrow} \hat n_{i\downarrow}, \label{eq:H_tJU}\\
\end{equation}
where: $\sum_{\left\langle i\, j \right\rangle}$ denotes the summation over the nearest neighboring sites, $t$ is the nearest-neighbor hopping integral, $J$ is the effective antiferromagnetic exchange integral, $\bS_i$ is the spin operator in the fermion representation, and $U$ is the on-site Coulomb repulsion magnitude.

One methodological remark is in place here. Usually, when starting from the Hubbard or $t$--$J$ models and discussing subsequently the correlated states and phases, one neglects the intersite repulsive Coulomb interaction $\sim\ K \sum_{\langle i j \rangle} \hat n_i \hat n_j$, where $\hat n_i = \sum_\sigma \hat n_{i\sigma}$ is the number of particles on site $i$. In the strong-correlation limit $U \gg W$ the corresponding transformation to the effective $t$-$J$ model provides\cite{Spalek2007-ActaPhysPolonA.111.409} the effective exchange integral $J_{ij} = 4 t_{ij}^2 / (U-K)$, and since $K \sim U/3$, we have a strong enhancement ($\sim 30\%$) of the kinetic exchange integral.
Strictly speaking, the contribution $\sim K$ should be then also added to the effective Hamiltonian \bref{eq:H_tJU}. However, this term has been neglected, as well as the similar contribution $\sim (J/4) \sum_{\langle i j \rangle} \hat n_i \hat n_j$ appearing in the full Dirac exchange operator \cite{Spalek2007-ActaPhysPolonA.111.409}, since we assume that the physically meaningful regime is that with $U \gtrsim W \gg K$ so that any charge-density wave instability is irrelevant in this limit.

We study properties of the above Hamiltonian using the Gutzwiller variational approach,\cite{Gutzwiller1963-prl, *Gutzwiller1965-pr} in which the trial wave function has the form\cite{Laughlin2006, Yuan2005-prb, Heiselberg2009-pra} $|\Psi\rangle = \hat P_G |\Psi_0\rangle$, where $\hat P_G$ is an operator specifying explicitly the configurations with double on-site occupancies, and $|\Psi_0\rangle$ is 
an eigenstate of a single-particle Hamiltonian (to be defined later).
 Since the correlated state $|\Psi\rangle$ is related to $|\Psi_0\rangle$, the average value of the Hamiltonian~$\mathcal{\mathcal{\hat H}}$ can be expressed as
\begin{equation}
 \frac{\langle\Psi|\mathcal{\hat H}|\Psi\rangle}{\langle\Psi|\Psi\rangle} =
\frac{\langle\Psi_0|\hat P_G \mathcal{\hat H} \hat P_G|\Psi_0\rangle}{\langle\Psi_0| \hat P_G^2 |\Psi_0\rangle}
\approx \langle \Psi_0 | \mathcal{\hat H}_{\itext{eff}} | \Psi_0 \rangle \equiv \langle\mathcal{\hat H}_{\itext{eff}}\rangle_0, \label{eq:rownowaznoscRMFT_GA}
\end{equation}
where $\langle \ldots \rangle_0$ means the average evaluated with respect to $| \Psi_0 \rangle$, and\cite{Zhang2003-prl, Gan2005-prb, Gan2005-prl, Yuan2005-prb, Heiselberg2009-pra}
\begin{equation}
   \mathcal{\mathcal{\hat H}}_{\mathit{eff}} = \ g_t t \sum_{\left\langle i\, j \right\rangle,\, \sigma} \left( \hat c^\dagger_{i\sigma} \hat c_{j \sigma} + \text{H.c.} \right) + g_s J \sum_{\left\langle i\, j \right\rangle} \bS_i \cdot \bS_{j} + U d^2 \label{eq:H_eff_tJU}\\
\end{equation}
is the effective Hamiltonian resulting from the Gutzwiller approximation\cite{Gutzwiller1963-prl, *Gutzwiller1965-pr} (GA). In the above formula,\ ${d^2 \equiv \left\langle \hat n_{i\uparrow} \hat n_{i\downarrow} \right\rangle_0}$ is the double-occupancy probability, $g_t$ and $g_s$ are the so-called Gutzwiller renormalization factors determined by the statistical counting of configuration with given $Nd^2$, $Nw$ and $Nr$.
(cf. Refs.~\onlinecite{Ogawa1975-ProgTheorPhys.53.614, Zhang1988-SuperSciTech.1.36}) 
\begin{subequations}
  \begin{eqnarray}
 g_t & = & \frac{n-2d^2}{n-2rw} \left( \sqrt{\frac{(1-w)(1-n+d^2)}{1-r}} \right. + \left. \sqrt{\frac{w d^2}{r}} \right) \nonumber\\
& & \times \left( \sqrt{\frac{(1-r)(1-n+d^2)}{1-w}} + \sqrt{\frac{rd^2}{w}} \right), \label{eq:tJU:gt}\\
 g_s & = & \left( \frac{n-2d^2}{n-2rw} \right)^2, \label{eq:tJU:gs}
\end{eqnarray}
\end{subequations}
where $n$ is the average number of electrons (occupancy) per site. To discuss AF order, the lattice is divided into two interpenetrating sublattices: $A$, where the majority of spins are oriented~$\uparrow$, and $B$, where the majority of spins are oriented~$\downarrow$. For sublattice $A$, $r \equiv \langle \hat n_{i\uparrow} \rangle = \frac{1}{2} \left( n + m_{AF} \right)$ and $w \equiv \langle \hat n_{i\downarrow} \rangle = \frac{1}{2} \left( n- m_{AF} \right)$, where $m_{AF}$ is the antiferromagnetic (staggered) spin polarization per site. For sublattice $B$, the definitions of $w$ and $r$ are interchanged.
Note, that the Gutzwiller factor \bref{eq:tJU:gs} has the same form for both $\frac{1}{2} \left( \hat S^x_i \hat S^y_j + \hat S^x_i \hat S^y_j \right)$ and $\hat S^z_i \hat S^z_j$ parts of $\bS_i \cdot \bS_{j}$. In a refined approach, two distinct Gutzwiller factors $g_s^{xy}$ and $g_x^z$ may be considered (cf. Ref.~\onlinecite{Himeda1999-PRB.60.R9935}). However, in this paper it is assumed that $g_s^{xy}=g_x^z \equiv g_s$, which is broadly accepted (see e.g. Refs.~\onlinecite{Zhang2003-prl, Gan2005-prb, Gan2005-prl, Yuan2005-prb, Heiselberg2009-pra}).
The reason is that the spin-singlet paired state is spin-rotationally invariant and in the case of coexistent antiferromagnetic state we limit ourselves to the mean-field-approach paradigm with the resulting N\'{e}el state.

\begin{figure}
 \centering
  \includegraphics[width=0.7\linewidth]{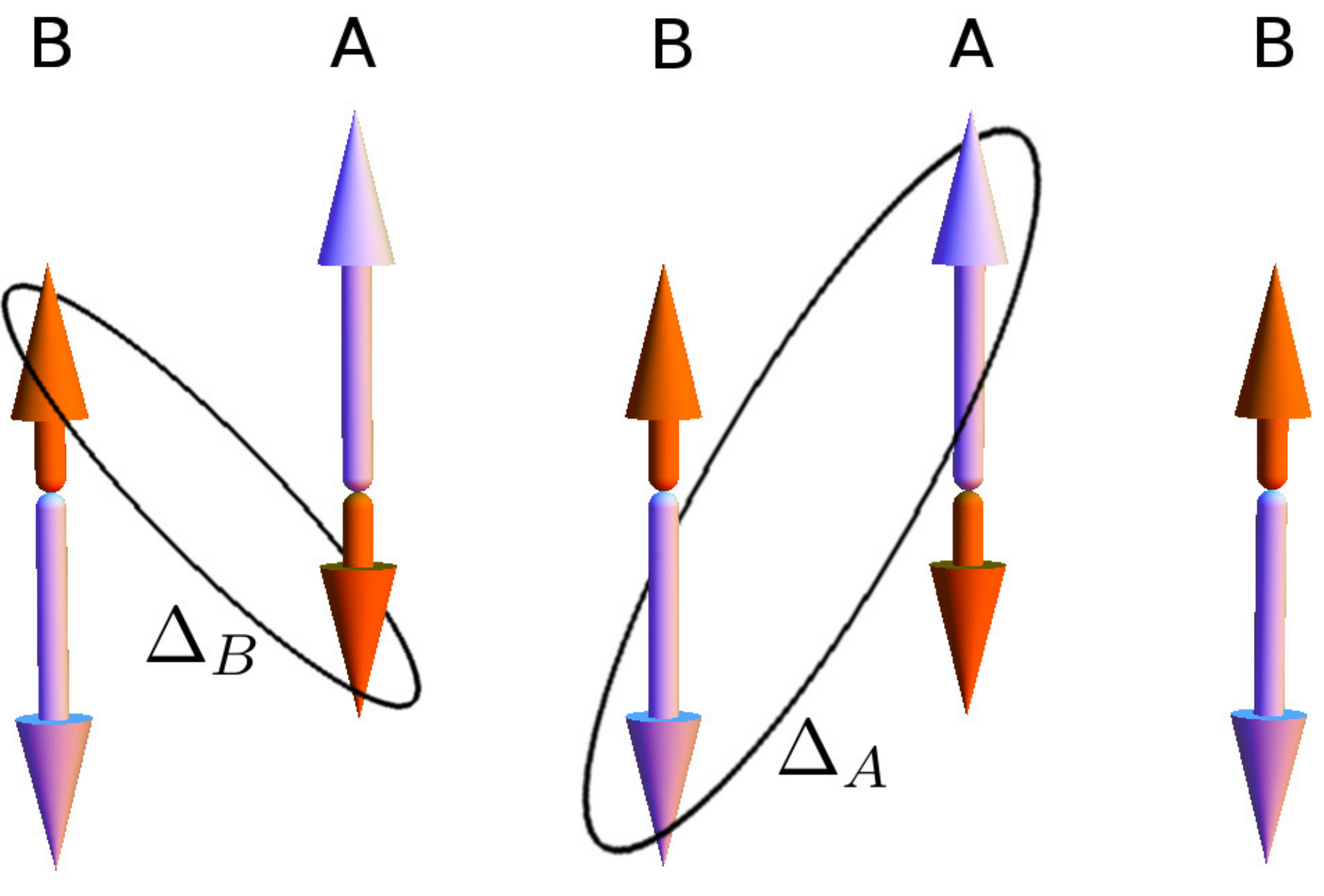}
 \caption{(Color online)
Schematic representation of the difference between the pairing parameters for the majority-spin electrons (large arrows) and the minority-spin electrons (small arrows) in the two-sublattice system with AF order.
Since for $m_{AF} \neq 0$ there may be that the real part of $\Delta_A$ and $\Delta_B$ might be different, and a spin-triplet component of the superconducting order has to be considered [cf. Eq.~\bref{eq:defDeltaT}].}
 \label{fig:2deltas}
\end{figure}

In order to evaluate $\langle\mathcal{\hat H}_{\itext{eff}}\rangle_0$ we define
the average number of electrons per site with spin $\sigma$ as
\begin{equation}
 n_{i\sigma} \equiv \langle \hat c^\dagger_{i \sigma} \hat c_{i \sigma} \rangle_0 = \frac{1}{2} \left( n + \sigma\, e^{i {\bf Q} \cdot {\bf R}_i}\, m_{AF} \right),
\label{eq:n_boxG}
\end{equation}
with ${\bf Q} \equiv (\pi,\, \pi)$, and with ${\bf R}_i$ denoting position vector of site $i$ and
the following bare (nonrenormalized) quantities:
the hopping amplitude for the nearest neighbors $\langle i, j \rangle$ as
\begin{equation}
   \chi_{ij\sigma} \equiv \langle \hat c^\dagger_{i \sigma} \hat c_{j \sigma} \rangle_0 = \chi_{AB},
\label{eq:chi_boxG}
\end{equation}
and the pairing order parameter in real space in the form
\begin{equation}
 \Delta_{ij\sigma} \equiv \langle \hat c_{i \sigma} \hat c_{j \bar\sigma} \rangle_0 \equiv - \tau_{ij}\left(\sigma \Delta_S + e^{i {\bf Q}\cdot {\bf R}_i} \Delta_T\right),
\label{eq:Delta_boxG}
\end{equation}
where $\tau_{ij} \equiv 1$ for $j=i\pm \hat x$ and ${\tau_{ij}\equiv -1}$ for $j=i\pm \hat y$ (in order to ensure the $d$-wave symmetry). In consequence, the spin-singlet $(\Delta_S)$ and the spin-triplet $(\Delta_T)$ components of the gap are defined as
\begin{subequations}
\begin{eqnarray}
 \tau_{ij}\Delta_S & = & \frac{1}{4} \left(
\left\langle \hat c_{j\in B\, \downarrow} \hat c_{i\in A \uparrow} \right\rangle_0 +
\left\langle \hat c_{i\in A\, \downarrow} \hat c_{j\in B \uparrow} \right\rangle_0 +
\mbox{H.c.} \right) \nonumber\\
& = & \frac{1}{4} \tau_{ij} \left( \Delta_A + \Delta_B + \mbox{H.c.} \right),
\label{eq:defDeltaS}\\
\tau_{ij}\Delta_T & = & \frac{1}{4} \left(
\left\langle \hat c_{j\in B\, \downarrow} \hat c_{i\in A \uparrow} \right\rangle_0 -
\left\langle \hat c_{i\in A\, \downarrow} \hat c_{j\in B \uparrow} \right\rangle_0 +
\mbox{H.c.} \right) \nonumber\\
& = & \frac{1}{4} \tau_{ij} \left( \Delta_A - \Delta_B + \mbox{H.c.} \right). \label{eq:defDeltaT}
\end{eqnarray}
\end{subequations}
In some works (e.g. in Refs.~\onlinecite{Yuan2005-prb, Heiselberg2009-pra}) the triplet component is disregarded.
However, since $\Delta_A$ represents an average pairing for {\it majority spins} on nearest neighboring sites and $\Delta_B$ an average pairing of {\it minority spins} (when AF order is present, cf. Fig.~\ref{fig:2deltas}), the real part of $\Delta_A$ and $\Delta_B$ might be different (cf. also the work of Tsonis\cite{Tsonis2008-jpcm} and Aperis\cite{Aperis2010-prl} regarding the inadequacy of a single-component order parameter to describe the SC phase). Therefore, in this paper, this more comprehensive  structure is introduced.
Nonetheless, in order to evaluate the significance of introducing the triplet term for the SC gap, the results are compared also with those obtained for the case when $\Delta_T$ is set to zero.

Applying the Wick's theorem to the Eq.~\bref{eq:rownowaznoscRMFT_GA}, the expectation value $\left\langle \mathcal{\mathcal{\hat H}}_{\mathit{eff}} \right\rangle_0 \equiv W$ can be obtained in the form (see Appendix A for details)
\begin{equation}
 \frac{W}{\Lambda} = 8 g_t t \chi_{AB}  + U d^2  - g_s J \left( \frac{1}{2} m_{AF}^2 +3 \chi_{AB}^2 + 3 \Delta_S^2 - \Delta_T^2 \right),
\label{eq:W}
\end{equation}
where $\Lambda$ is the number of atomic sites in the system.
Note that the total energy of this correlated system is composed of three interdependent parts: (i) the renormalized hopping energy $\sim t g_t \chi_{AB} <0$, (ii) the correlation energy $Ud^2 > 0$, and (iii) the exchange contribution $\sim g_s J$ lowering both the energies of AF and SC states. This balance of physical energies will be amended next by the constraints introducing the statistical consistency into this mean-field system to guarantee that the self-consistent and the variational procedures will lead to the same single-particle states (this is the so-called Bogoliubov principle for the optimal single-particle states))

To summarize, the process of derivation of the effective single-particle Hamiltonian \bref{eq:H_eff_tJU} is fully justified by its definition \bref{eq:rownowaznoscRMFT_GA} which involves an averaging procedure over an uncorrelated state $| \Psi_0 \rangle$. This state is selected implicitly. In general, it is the state with broken symmetry, i.e. with nonzero values of $m_{AF}$, $\Delta_S$, and $\Delta_T$.
In other words, $| \Psi_0 \rangle$ is defined through the values of order parameters to be determined either self-consistently or variationally. This is the usual procedure proposed originally by Bogoliubov\cite{Bogoliubov1958, *Bogoliubov1958-USSR, *BogoliubovTolmachev1958} in his version of BCS theory and by Slater\cite{Slater1951-PhysRev.82.538} in the theory of itinerant antiferromagnetism. Here their simple version of mean theory becomes more sophisticated, since the renormalization factors contain also the order parameters and in a singular formal form. This last feature leads to basic formal changes in formulation of the renormalized mean-field theory, as discussed next.

\section{Quasiparticle states and minimization procedure for the ground state}

Following Refs.~\onlinecite{Jedrak2010-prb, Jedrak2011-PhD, Jedrak2010-arXiv, Kaczmarczyk2011-prb, Kaczmarczyk2011-PhD}, we write the mean-field grand Hamiltonian in the form
\begin{multline}
\allowdisplaybreaks
 \hat K \equiv  W - \sum_{\left\langle i\, j \right\rangle,\,\sigma} \left( \lambda_{ij\sigma}^\chi \left( \hat c_{i\sigma}^\dagger \hat c_{j\sigma} - \chi_{ij\sigma} \right) + \text{H.c.} \right)\\
- \sum_{\left\langle i\, j \right\rangle,\,\sigma} \left( \lambda_{ij\sigma}^\Delta \left( \hat c_{i\sigma} \hat c_{j\bar\sigma} - \Delta_{ij\sigma} \right) + \text{H.c.} \right) \\
- \sum_{i\sigma} \left( \lambda_{i\sigma}^n \left( \hat n_{i\sigma} - n_{i\sigma} \right) \right)
- \mu \sum_{i\sigma} \hat n_{i\sigma}, \label{eq:K}
\end{multline}
where $\mu$ is the chemical potential, and the Lagrange multipliers $\{ \lambda \}$ are introduced for each operator whose average appears in $W$ [Eq.~\bref{eq:W}]. The Lagrange multipliers can be interpreted as the correlation-induced effective fields.
We should underline, that the additional terms guarantee that the averages calculated in a self-consistent manner coincide with those determined from variational minimization principle of the appropriate free- or ground-state energy functional.
Due to the dependence of the renormalization factors on the mean-field values, the two ways of their calculation do differ, but the introduced constraints ensure their equality. In this manner, as said above, the approach is explicitly in agreement with the Bogoliubov theorem that the single-particle approach represents the optimal formulation from the principle of maximal-entropy point of view\cite{Jedrak2010-prb, Jedrak2011-PhD}. Also,
the fields $\{ \lambda \}$ are assumed to have the same symmetry as the broken-symmetry states, to which they are applied [cf. Eqs.~\bref{eq:n_boxG}, \bref{eq:chi_boxG}, and \bref{eq:Delta_boxG}]. Namely,
\begin{subequations}
 \begin{eqnarray}
 \lambda^n_{i\sigma} & = & \frac{1}{2} \left( \lambda_n + \sigma  e^{i {\bf Q} \cdot {\bf r}_i} \lambda_{m} \right), \label{eq:lambda-first}\\
  \lambda^\chi_{ij\sigma} & = &  \lambda_{\chi}, \label{eq:lambda-middle}\\
  \lambda^{\Delta}_{ij\sigma} & = & - \tau_{ij} \left( \sigma \lambda_{\Delta_S} + e^{i {\bf Q} \cdot {\bf r}_i} \lambda_{\Delta_T} \right). \label{eq:lambda-last}
 \end{eqnarray}
\end{subequations}

To solve Hamiltonian \bref{eq:K}, space Fourier transformation is performed first. Then, the Hamiltonian is diagonalized and yields four branches of eigenvalues (details of the calculations are presented in Appendix \ref{app:HowToGetTheEquations}).
Next, we define the generalized grand potential functional at temperature $T>0$ as given by
 \begin{equation}
 \mathcal{F} = -\frac{1}{\beta} \ln{\mathcal{Z}}, \hspace{6pt} \mbox{with} \hspace{6pt}  \mathcal{Z} = \mathrm{Tr}\left( e^{-\beta \hat K} \right), \label{eq:Funkcjonal_SGA_def}
\end{equation}
with $\beta \equiv 1/k_B T$.
Explicitly, $\mathcal{F}$ then has the following form [cf. Eq.~\bref{eq:Funkcjonal_SGA-pelny2} in Appendix \ref{app:HowToGetTheEquations}]:
\begin{multline}
 \mathcal{F} / \Lambda= 
8 g_t t \chi_{AB} 
- g_s J \left( \frac{1}{2} m_{AF}^2 + 3 \chi_{AB}^2 + 3 \Delta_S^2 - \Delta_T^2 \right) \\
+ \frac{1}{2} \lambda_n(n-1) + \frac{1}{2} \lambda_{m} m_{AF}
+ 8 \left( \lambda_{\chi} \chi_{AB} + \lambda_{\Delta_S} \Delta_S + \lambda_{\Delta_T} \Delta_T \right)\\
-\frac{1}{\Lambda \beta} \sum_{l,\bk} \ln(1 + e^{-\beta E_{l\bk}})
+ U d^2 - \mu.
\label{eq:Funkcjonal_SGA-pelny}
\end{multline}
The necessary conditions for the minimum of $\mathcal{F}$ subject to all constraints 
are
\begin{equation}
 \frac{\partial \mathcal{F}}{\partial \vec A}=0, \hspace{6pt}  \hspace{6pt} \frac{\partial \mathcal{F}}{\partial \vec \lambda}=0, \hspace{6pt} \mbox{and} \hspace{6pt} \frac{\partial \mathcal{F}}{\partial d}=0,\label{eq:dFdl_dFdA}
\end{equation}
where the five mean-field parameters are labeled collectively as $\vec A$, and the Lagrange multipliers as $\vec \lambda$ [the full form of Eqs.~\bref{eq:dFdl_dFdA} is presented in Appendix~\ref{app:Equations-explicite}].
Note that five above equations ($\partial \mathcal{F}/\partial \vec A=0$) can be easily eliminated, reducing the system of algebraic equations to be solved (cf. Appendix~\ref{app:Equations-explicite} and discussion in Appendix~\ref{app:2-schemas-discussion}).

One should note one nontrivial methodological feature of the approach contained in the grand Hamiltonian \bref{eq:K}. namely, the effective Hamiltonian \bref{eq:H_eff_tJU} appears in it in the form of expectation value $W$ [cf. Eq.~\bref{eq:W}], whereas the constraints appear in Eq.~\bref{eq:K} in the explicite operator form. This is a nonstandard mean-field version of approach. The correspondence to and main difference with the standard renormalized mean-field approach is discussed in Appendix~\ref{app:2-schemas-discussion}.

As we are interested in the ground-state properties (${T=0}$), we take the ${T \rightarrow 0}$ limit. We have checked that taking $\beta^{-1} = k_B T = 0.002\, |t|$ is sufficient for  practical purposes.\footnote{With increase of hole-doping the approximation of zero temperature described in the main text became weaker. For bigger $\delta$ (e.q. for ${U/|t|=12}$ bigger than $0.3$) it starts be insufficient. Taking bigger $\beta$ moves the limiting value of $\delta$ just a little. Thus, for limit of strong hole-doping other computational techniques have to be used (since the correlations are weak in such limit, it can be a basic RMFT methods as described in Ref.~\protect\onlinecite{Yuan2005-prb}).}

\section{Results: Phase diagram and microscopic characteristics}
\label{sec:results}

The stable phase is determined by the solution which has the lowest physical free energy defined as minimal value of
\begin{equation}
F =  \mathcal{F}_0 +  \mu \Lambda n, \label{eq:FreeEnergy}
\end{equation}
where $\mathcal{F}_0$ denotes the value of $\mathcal{F}$ obtained at the minimum [cf. conditions \bref{eq:dFdl_dFdA}].

The minimum of  $\mathcal{F}$ was obtained numerically using \texttt{GNU Scientific Library (GSL)},\cite{GSL-manual} and unless stated otherwise,  all calculations were made for $t=-1$, $J=|t|/3$, $\beta|t|=500$ on a two-dimensional square lattice of size $\Lambda=512\times 512$ with periodic boundary conditions.

\begin{figure}
  \centering
  \includegraphics[width=1\linewidth]{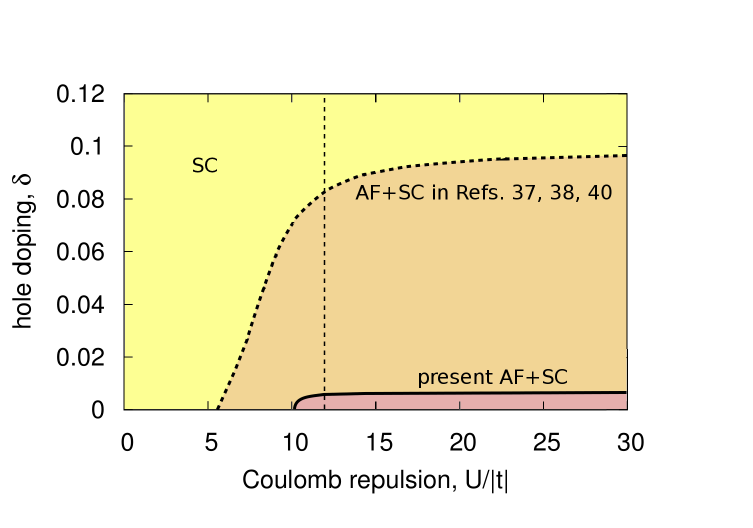}

  \caption{(Color online) Representative phase diagram for the \tJU\ model on the Coulomb repulsion -- hole doping plane. The phases are labeled as follows: SC -- superconducting phase, AF+SC -- phase with coexisting superconducting and antiferromagnetic orders. The pure stable AF phase is found only for $\delta=0$ and for $U>10.6|t|$. The value $U/|t|=12$ marked by the dashed vertical line is taken in the subsequent analysis.
  The solid line is a our result. The dashed line is the result of previous studies (Refs.~\onlinecite{Yuan2005-prb, Heiselberg2009-pra, Fen2012-CommTheorPhys}).}
  \label{pic:phaseDiagram-tJU}
\end{figure}

A representative phase diagram on the Coulomb repulsion $U$ -- hole doping $\delta$ plane is exhibited in Fig.~\ref{pic:phaseDiagram-tJU}. We find three stable phases: SC, AF and phase with coexisting SC and AF order (labeled collectively as AF+SC). The pure AF stable phase is found only for $\delta \equiv 1-n = 0$ and  $U/|t|>10.6$. The region where the AF+SC appears is limited to a very close proximity to the Mott insulating state (hole-doping range ${\delta \in (0,\, 0.006)}$).
%
Our results differ significantly from previous studies (cf., e.q., Refs.~\onlinecite{Yuan2005-prb, Heiselberg2009-pra, Fen2012-CommTheorPhys}), where a much wider coexistence region was reported (dashed line in Fig.~\ref{pic:phaseDiagram-tJU}). The previous results were an effect of the non-statistically-consisted RMFT approach used, as also is explained below. Using our method, such a consistency is achieved, and as a result a much narrower coexistence regime appears. It squares with recent experimental studies, where the region of AF+SC was reported to be narrow \{cf. e.g. Bernhard\cite{Bernhard2012-PhysRevB.86.184509} [study of Ba(Fe$_{1-x}$Co$_x$)$_2$As$_2$], where the coexistence region is not wider than $0.02$ (of the hole doping range)\}.

\begin{figure}
  \centering
  \includegraphics[width=1\linewidth]{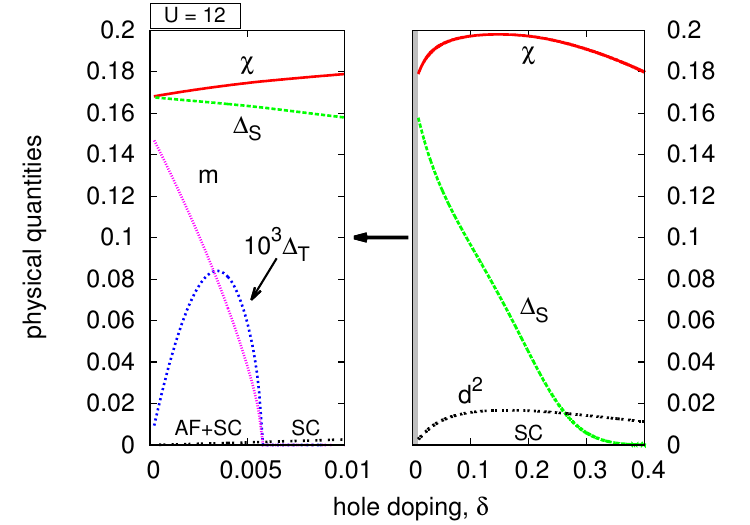}

  \caption{(Color online) 
   Selected bare physical quantities $\chi_{AB}$, $\Delta_S$, $\Delta_T$, $m_{AF}$, and $d^2$, all as a function of doping $\delta$ and for $U/|t|=12$.
In the coexistent AF+SC phase we have $\Delta_S \neq 0$, $\Delta_T \neq 0$, and $m_{AF} \neq 0$.
Such phase is stable only in the vicinity of the half-filling, as detailed in the left panel where $\delta \in [0,\, 0.01]$.
On the right panel we present overall behavior in the wider range of the doping.).
} \label{pic:phase-diagram-cut-U12-1}
 \end{figure}

\begin{figure}
  \includegraphics[width=1\linewidth]{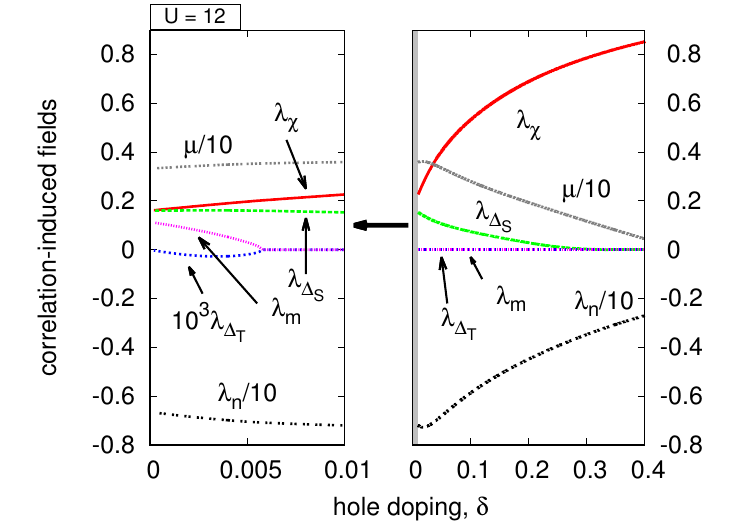}

  \caption{(Color online) Correlation-induced fields $\lambda{\chi_{AB}}$, $\lambda_{\Delta_S}$, $\lambda_{\Delta_T}$, $\lambda_{m}$, and the chemical potential $\mu$, all as a function of doping $\delta$ for $U/|t|=12$. The fields $\{\lambda \}$ have non-zero values only when the corresponding mean fields are also non-zero.
Note that on the left panel $\mu \sim W/2 \sim 4|t|$, corresponding to the almost-filled lower Hubbard subband.
} \label{pic:phase-diagram-cut-U12-2}
 \end{figure}

For further analysis we restrict ourselves to $U/|t|=12$, as marked by the dashed vertical line in Fig.~\ref{pic:phaseDiagram-tJU}. In Figs.~\ref{pic:phase-diagram-cut-U12-1} and~\ref{pic:phase-diagram-cut-U12-2}, we plot the doping dependence of the mean-fields and the correlation fields.
The magnitude of $\Delta_T$ is non-zero only in the region with AF order (i.e. when ${m_{AF} \neq 0}$).

The correlated spin-singlet gap parameter in real space is defined as
\begin{eqnarray}
 \tau_{ij}\Delta^c_S & = & \frac{1}{4}  \left(
\left\langle \hat c_{i\in A\, \downarrow} \hat c_{j\in B \uparrow} \right\rangle -
\left\langle \hat c_{i\in A \uparrow} \hat c_{j\in B\, \downarrow} \right\rangle +
\mbox{H.c.}
\right), \label{eq:defDeltaS_r}
\end{eqnarray}
where the average is calculated using the Gutzwiller wave function $\left| \Psi \right\rangle$, instead of $\left| \Psi_0 \right\rangle$. 
Approximately (within GA), the correlated (physical) SC order parameters can be expressed as\cite{Yuan2005-prb, Heiselberg2009-pra}
\begin{eqnarray}
\Delta^{c}_S & = & g_\Delta \Delta_S, \hspace{12pt}\mbox{and}\hspace{12pt}
\Delta^{c}_T = g_\Delta \Delta_T,
\end{eqnarray}
where 
\begin{eqnarray}
   g_\Delta & = &
\frac{n-2d^2}{2(n-2rw)} \left[ \left( \sqrt{\frac{(1-w)(1-n+d^2)}{1-r}} + \sqrt{\frac{w d^2}{r}} \right)^2 \right. + \nonumber\\
& & + \left. \left( \sqrt{\frac{(1-r)(1-n+d^2)}{1-w}} + \sqrt{\frac{rd^2}{w}} \right)^2 \right]. \label{eq:tJU:gD}
\end{eqnarray}
The AF order parameter, and the renormalized hopping parameter
are defined in a similar manner, specifically
\begin{eqnarray}
m^{c}_{AF} & = & g_m\, m_{AF}, \\
\chi^{c}_{AB} & = & g_t\, \chi_{AB},
\end{eqnarray}
where $g_t$ is presented in Eq.~\bref{eq:tJU:gt} and
\begin{eqnarray}
g_m & = &  \frac{n-2d^2}{n-2wr}. \label{eq:tJU:gm}  
\end{eqnarray}

 \begin{figure}
  \centering
  \includegraphics[width=1\linewidth]{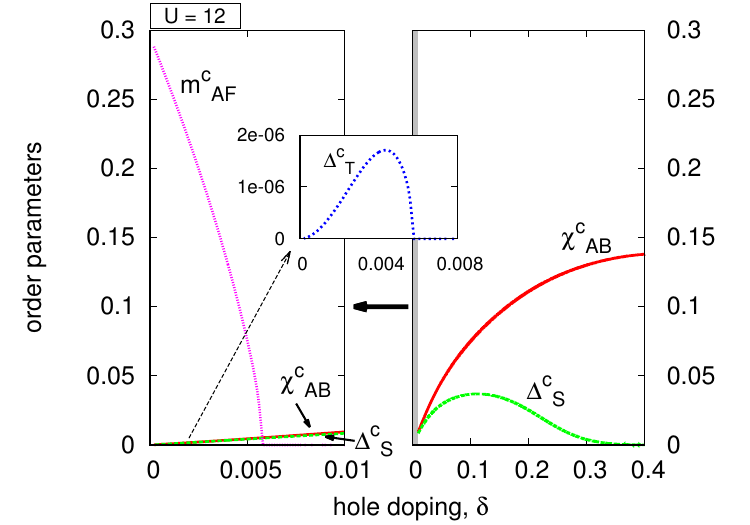}

  \caption{(Color online) Order parameters of SC and AF states in the correlated state versus doping $\delta$ and for $U/|t|=12$. AF disappears for $\delta \approx 6 \cdot 10^{-3}$. In the limit $\delta = 0$ the state transforms into an AF insulator. For $\delta \gtrsim 0.1$, the hopping amplitude $\chi^c_{AB}$ increases substantially.
  Inset: dependence of $\Delta^c_T$ in the vicinity of the half-filling. Note, that the value of $\Delta^c_T$ is about $10^4$ times smaller than the value of $\Delta^c_S$.
} \label{pic:phase-diagram-cut-U12-3}
 \end{figure}

 \begin{figure}
  \centering
  \includegraphics[width=1\linewidth]{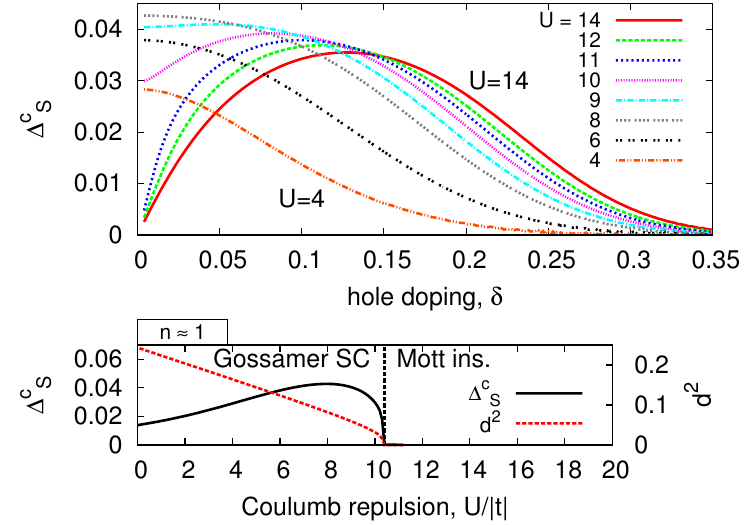}
  \caption{(Color online) Profile of the correlated singlet gap $\Delta^c_{S}$ for selected values of $U/|t|$ versus hole-doping (top). The limiting values of $\Delta^c_{S}$ and $d^2$ for $n \approx 1$ are presented in the bottom panel.} \label{pic:profile-gossamer}
 \end{figure}

The magnitude of $\Delta_T^c$ is about $10^4$ times smaller than the magnitude of $\Delta_S^c$, so most probably, it may not be observable. For $\delta \gtrsim 1/3$ the order parameter $\Delta_S^c$ decrease exponentially.
Note a spectacular increase of the hopping probability $\chi^c_{AB}$ with increased doping in Fig.~\ref{pic:phase-diagram-cut-U12-3}, leading to an effective Fermi liquid state for $\delta \gtrsim 1/3$.

The non-zero correlated gap at $n=1$ for low-$U$ values provides an evidence for a gossamer superconductivity.
The concept of gossamer superconductivity was introduced by Laughlin\cite{Laughlin2006} and it describes the situation when the pure SC phase is stable at the half-filling.
For $U/|t| \approx 10.6$ and $n=1$, where AF+SC phase sets in, the correlated gap $\Delta^{c}_S$ vanishes. Details of the transition are presented in Fig.~\ref{pic:profile-gossamer}, cf. the bottom panel. The critical $U/|t|$ value for the disappearance of $\Delta_S$ is marked by the dotted vertical line.

 \begin{figure}
  \includegraphics[width=1\linewidth]{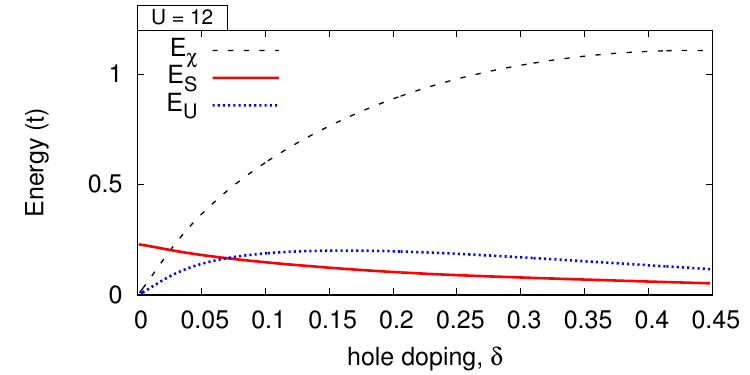}

  \caption{(Color online) The optimized component energies composing the total energy (Eq. \bref{eq:W}, which represent $T \rightarrow 0$ limit of Eq.~\bref{eq:Funkcjonal_SGA-pelny}). For discussion see main text.} \label{pic:porownanie-energii}
 \end{figure}

In the sake of completeness, we have drawn for $U=12|t|$ in Fig.~\ref{pic:porownanie-energii} the components of the total energy (Eq.~\bref{eq:Funkcjonal_SGA-pelny}), to show that in the underdoped regime the effective hopping energy $E_\chi \equiv 8 g_t |t| \chi_{AB}$, the total exchange contribution $E_S \equiv g_s J \left( m_{AF}^2/2  + 3 \chi^2_{AB} + 3 \Delta^2_S - \Delta^2_T \right)$, and the Coulomb energy $E_U = U d^2$, are all of comparable magnitude. This is the regime of strong correlations.

The overall behavior of the obtained characteristics can be summarized as follows. First, the coexistent AF+SC phase appears only for the doping $\delta<0.006$ and transforms into the pure Mott insulating state (AF) only at the half filling $\delta=0$ and large $U$. The spin-triplet gap component is practically negligible in the AF+SC phase.
Introducing the molecular fields $\lambda_\chi$, $\lambda_{\Delta_S}$, and $\lambda_m$ (where non-zero), and $\lambda_n$ change the phase diagram in a significant manner which means, that the influence of the consistency constraints on the single-particle states is important.
The spin-singlet $d$-wave superconductivity vanishes exponentially for large $\delta$. The optimal doping appears in the interval $\delta \sim 0.1$ -- $0.15$ and is weakly dependent on $U$ for $U \gtrsim 12|t|$.
%
%

\section{Conclusions and comments}
\label{sec:conclusionANDoutlook}

Using the statistically-consistent Gutzwiller approximation (SGA), we have analyzed in detail the effective Hamiltonian considered previously in Refs.~\onlinecite{Yuan2005-prb, Heiselberg2009-pra, Fen2012-CommTheorPhys}.
However, in contrast to those papers, we have considered a more complete structure of the SC gap (the components $\Delta_S$ and $\Delta_T$).
Also, a significantly narrower region of the coexistence of AF and SC is obtained. Furthermore, the critical value of $U$ for AF+SC appearance is higher, and for $J/|t| = 1/3$ the value is about $10.6|t|$.
We have checked that the bare amplitude $\Delta_T$ is about $10^3$ times smaller than that of $\Delta_S$ (similarly, the order parameters ratio $\Delta_T^c/\Delta_S^c \approx 10^{-4}$).
We have checked that when the $\Delta_T$ is omitted, the results do not change in any significant manner. Therefore, the spin-triplet component of the superconducting order is most probably not detectable experimentally.

In previous studies (cf. Refs.~\onlinecite{Yuan2005-prb, Heiselberg2009-pra, Fen2012-CommTheorPhys}) a much wider coexistence region was reported. In the present paper, we correct those predictions (cf. Fig. \ref{pic:phaseDiagram-tJU}). Namely, we show, that the previous results were an effect of the non-statistically consistent RMFT approach used.
Illustratively, in Ref.~\onlinecite{Yuan2005-prb}, a minimization procedure is formulated by setting $\partial E_{\mbox{var}}/\partial m=0$, yielding Eq.~(21) in Ref.~\onlinecite{Yuan2005-prb} for $m$ which is different than that defining $m$ [cf. Eq.~(16) in Ref.~\onlinecite{Yuan2005-prb}].
We claim that a more correct approach is provided by SGA, where the Lagrange multipliers are introduced for each operator for which the average appears in the effective mean-field Hamiltonian.
In other words, without incorporating the multipliers, the free energy functional $\mathcal{F}$ is minimized in an over-extended Fock space containing, along with physical configurations, also those that lead to the statistical inconsistency. Using the constraints introduced by SGA, this space is limited to a subspace, in which such an inconsistency does not appear.
Hence, the energy obtained in SGA can either be equal to or even {\it higher} than the energy obtained using non-consistent approaches. Obviously, this circumstance should not be used as an argument against SGA.
Different formulations, where the model is solved in a self-consistent manner are also presented in Refs.~\onlinecite{Bunemann-jp, Kai-Yu-njp}.

As said above, in the SGA method, an effective single-particle approach with conditions \bref{eq:dFdl_dFdA} is developed. In such an approach the question of a pseudogap is not addressed. This is because (i) the order parameter $\Delta_S$ is assumed as real (i.e., no phase fluctuation appears), and (ii) the collective spin degrees of freedom are not separated from single-particle fermionic correlations.
In order to address that issue, one would have to generalize the approach to include, e.g., the spin sector of the excitations,\cite{Feng2012-PRB.85.054509} even in the absence of AF order.
As the antiferromagnetism is built into the SGA approach automatically, work on extension of this approach to include magnetic fluctuations in the paramagnetic phase is in progress.

One should note that the definition of the Mott (or Mott-Hubbard) insulator here complements that for the Hubbard model within the standard Gutzwiller approximation (GA) which represents the infinite-dimension variant of the approach.\cite{Kubo1975-ProgTheorPhys.54.1289} Namely, with an assumption that $J \neq 0$ we have a gradual evolution of the antiferromagnetic order parameter $m_{AF} \rightarrow 1$ with the increasing $U$, i.e., the system evolves from the Slater to the Mott antiferromagnet. This is what is also obtained in the saddle-point approximation within the slave-boson approach,\cite{Korbel2003} which differs from the standard GA by incorporating constraints, some of them of similar character as those introduced here in SGA.
In this respect both SGA and the saddle-point approximation to slave-boson approach go beyond GA, albeit not in an explicitly systematic formal manner.

\section*{Acknowledgments}

This research was supported in part by the Foundation for Polish Science (FNP) under
the Grant TEAM and in part by the National Science Center (NCN) through the Grant MAESTRO, No. DEC-2012/04/A/ST3/00342.


\appendix


\section{Definitions of the mean-fields and evaluation of $\left\langle \mathcal{\mathcal{\hat H}}_{\mathit{eff}} \right\rangle_0 \equiv W$}
\label{apen_W}

In the main text, the uniform bond order parameter for $i$ and $j$ sites indicating the nearest neighboring sites is defined as $\langle \hat c^\dagger_{i \sigma} \hat c_{j \sigma} \rangle_0 \equiv \chi_{AB}$. It was assumed that $\langle \hat c^\dagger_{i \sigma} \hat c_{j \sigma} \rangle_0$ is real. Let us consider in this appendix a more general form. Since the sublattice $A$ contains the sites where the majority spin is $\uparrow$ and the sublattice $B$ the sites where majority spin is $\downarrow$, the general form can be written as
\begin{equation}
 \langle \hat c_{i\in A\uparrow}^\dagger \hat c_{j \in B\uparrow} \rangle_0 = \langle \hat c_{j\in B\downarrow}^\dagger \hat c_{i\in A\downarrow}  \rangle_0 =
\langle \hat c_{i\in A\downarrow}^\dagger \hat c_{j\in B\downarrow} \rangle_0^* = \chi_{o-c},
\label{eq:def-chi1}
\end{equation}
\begin{equation}
\langle \hat c_{i\in A\downarrow}^\dagger \hat c_{j\in B\downarrow}  \rangle_0 =
 \langle \hat c_{j\in B\uparrow}^\dagger \hat c_{i\in A\uparrow} \rangle_0 =
\langle \hat c_{i\in A\uparrow}^\dagger \hat c_{j\in B\uparrow} \rangle_0^* = \chi_{c-o},
\end{equation}
for $i$ and $j$ being the nearest neighbors, where $\chi_{o-c}$ is an average of the operator describing the hopping of an electron from a site, where the average spin is {\it opposite} to the spin of the electron, to the site, where the average spin is {\it congruent} to the spin of the electron. $\chi_{o-c}$ describes the opposite situation. This results in the general expression that
\begin{equation}
   \chi_{ij\sigma} \equiv \langle \hat c^\dagger_{i \sigma} \hat c_{j \sigma} \rangle_0 = \chi_{AB} + i\sigma e^{i {\bf Q}\cdot {\bf R}_i}  \delta\chi_{AB},
\label{eq:chi_boxAp}
\end{equation}
where $\chi_{AB} \equiv \frac{1}{2} \left( \chi_{o-c} + \chi_{c-o} \right)$ and $\delta\chi_{AB} \equiv \frac{1}{2i} \left( \chi_{o-c} - \chi_{c-o} \right)$.

The electron-pairing order parameter for the nearest neighbors is defined as
\begin{equation}
 \Delta_{ij \downarrow} \equiv \left\langle \hat c_{j\, \downarrow} \hat c_{i \uparrow} \right\rangle_0 \equiv \left\{
   \begin{array}{ll}
     \tau_{ij} \tilde\Delta_{A}, & \text{for $i \in A$}, \\
     \tau_{ij} \tilde\Delta_{B}, & \text{for $i \in B$},
   \end{array}
\right. \label{eq:def-Delta}
\end{equation}
where $\tau_{ij} \equiv 1$ for $j=i\pm \hat x$ and ${\tau_{ij}\equiv -1}$ for $j=i\pm \hat y$ ($\Delta_{ij \uparrow}$ is defined in simillar manner). 
For the staggered magnetic moment $m_{AF}=0$ one can assume that $\tilde\Delta_{A} = \tilde\Delta_{B}$. However, when $m_{AF} \neq 0$, the order parameter $\tilde\Delta_{A}$ is a product of two operators, both of which annihilate electrons whose spin is {\it congruent} to the average spin of individual sites. On the contrary, $\tilde\Delta_{B}$ is a product of two operators that annihilate electrons whose spin is {\it opposite} to the average spin of individual sites.
Hence, it may be that $\tilde\Delta_{A} \neq \tilde\Delta_{B}$. Also, similar as with the hopping amplitude, $\tilde\Delta_{A}$ and $\tilde\Delta_{B}$ might be complex numbers. Let us denote $\tilde\Delta_A \equiv (\Delta_A,\ \delta\Delta_A)$ and $\tilde\Delta_B \equiv (\Delta_B,\ \delta\Delta_B)$, where the parameters in brackets are the real and imaginary parts of the corresponding gaps, respectively.

The only nontrivial part of $\left\langle \mathcal{\mathcal{\hat H}}_{\mathit{eff}} \right\rangle_0$ [cf. Eq.~\bref{eq:H_eff_tJU}] 
can be evaluated in the form
\allowdisplaybreaks
\begin{multline}
 4 \langle \bS_{i} \cdot \bS_{j} \rangle_0 \approx 
 -\left( \left\langle \hat c_{i\downarrow}^\dagger \hat c_{i\downarrow} \right\rangle_0 - \left\langle \hat c_{i\uparrow}^\dagger \hat c_{i\uparrow} \right\rangle_0 \right)^2 \\
- \left( \left\langle \hat c_{i\downarrow}^\dagger \hat c_{j\downarrow} \right\rangle_0 + 2 \left\langle \hat c_{i\uparrow}^\dagger \hat c_{j\uparrow} \right\rangle_0 \right) \left\langle \hat c_{j\downarrow}^\dagger \hat c_{i\downarrow} \right\rangle_0 \\
- \left( 2 \left\langle \hat c_{i\downarrow}^\dagger \hat c_{j\downarrow} \right\rangle_0 + \left\langle \hat c_{i\uparrow}^\dagger \hat c_{j\uparrow} \right\rangle_0 \right) \left\langle \hat c_{j\uparrow}^\dagger \hat c_{i\uparrow} \right\rangle_0 \\
- \left( -\left\langle \hat c_{i\uparrow} \hat c_{j\downarrow} \right\rangle_0 + 2 \left\langle \hat c_{i\downarrow} \hat c_{j\uparrow} \right\rangle_0 \right) \left\langle \hat c_{i\uparrow}^\dagger \hat c^\dagger_{j\downarrow} \right\rangle_0 \\
- \left( 2\left\langle \hat c_{i\uparrow} \hat c_{j\downarrow} \right\rangle_0 - \left\langle \hat c_{i\downarrow} \hat c_{j\uparrow} \right\rangle_0 \right) \left\langle \hat c_{i\downarrow}^\dagger \hat c^\dagger_{j\uparrow} \right\rangle_0, \label{eq:SS_average}
\end{multline}
where we have applied the Wick's theorem and we have assumed that $\left\langle \hat c_{i\uparrow}^\dagger \hat c_{i\downarrow} \right\rangle_0 \equiv 0$,
$\left\langle \hat c_{i\uparrow} \hat c_{i\downarrow} \right\rangle_0 \equiv 0$,
 and $\left\langle \hat c_{i\downarrow} \hat c_{j\downarrow} \right\rangle_0 = \left\langle \hat c_{i\uparrow} \hat c_{j\uparrow} \right\rangle_0 \equiv
0$.
Using the notation introduced above and Eq.~\bref{eq:n_boxG}, we have
\begin{multline}
 4\langle \bS_{i} \cdot \bS_{j} \rangle_0 = -m_{AF}^2 - 6\chi_{AB}^2 + 2(\delta\chi_{AB})^2 \\
- |\tilde\Delta_A|^2 - |\tilde\Delta_B|^2- 4\tilde\Delta_A \tilde\Delta_B.
\end{multline}
Since the above expression is invariant with respect to the same rotations of both vectors $\tilde\Delta_{A}$ and $\tilde\Delta_{B}$, one component of the vectors can be assumed to be eliminated. With the choice $\delta\Delta_A = 0$, we have    
\begin{multline}
 \Delta_{ij\sigma} \equiv \langle \hat c_{i \sigma} \hat c_{j \bar\sigma} \rangle_0
 \equiv - \tau_{ij}\left(\sigma \Delta_S + e^{i {\bf Q}\cdot {\bf r}_i} \Delta_T\right) \\
- \tau_{ij} \frac{1}{2} i \left(\sigma - e^{i {\bf Q}\cdot {\bf r}_i} \right)\delta\Delta_B,
\label{eq:Delta_boxAp}
\end{multline}
where $\Delta_S \equiv \Delta_A + \Delta_B$ and $\Delta_T \equiv \Delta_A - \Delta_B$.

Therefore, the $\left\langle \mathcal{\mathcal{\hat H}}_{\mathit{eff}} \right\rangle_0 \equiv W$ can be presented in the full form
\begin{multline}
 \frac{W}{\Lambda} = 8 g_t t \chi_{AB}  + U d^2 - g_s J \left( \frac{1}{2} m_{AF}^2 +3 \chi_{AB}^2 - (\delta\chi_{AB})^2 \right. \\
\left. + 3 \Delta_S^2 - \Delta_T^2 +\frac{1}{2} (\delta\Delta_B)^2 \right).
\label{eq:W-extraTerms}
\end{multline}

Introduction of $\delta\chi_{AB}$ and $\delta\Delta_{B}$ affects the form of selecting the correlated fields $\lambda^\chi_{ij\sigma}$ and $\lambda^{\Delta}_{ij\sigma}$, and the final set of necessary conditions for a local minimum of the free energy (cf. Eqs.~\bref{eq:lambda-middle}, \bref{eq:lambda-last} and \bref{eq:dFdl_dFdA}). However, it was found that the state with the lowest energy (for the considered model) has always been that with $\delta\chi_{AB} \equiv 0$ and $\delta\Delta_{B} \equiv 0$. Hence, it is acceptable to neglect both terms and clame that $\Delta_{ij\sigma}$ and $\chi_{ij\sigma}$ are both real. For simplicity and clarity it is how the averages are presented in the main text. Finally, Eq.~(\ref{eq:W-extraTerms}) is reduced to Eq.~(\ref{eq:W}).


\section{Determination of the grand potential functional (Eq.~\bref{eq:Funkcjonal_SGA-pelny})}
\label{app:HowToGetTheEquations}

To diagonalize $\hat K$ [Eq.~\bref{eq:K}], we first perform the space Fourier transform. The result can be rewritten in the following $4 \times 4$ matrix form

\vspace{-18pt}
\begin{multline}
 \hat K  = W + \sum_\bk{}^{'} \hat\Psi_\bk^\dagger \mathbbm{\tilde M}_\bk \hat\Psi_\bk
+ \frac{1}{2} \Lambda \left( \lambda_n (n-1) + \lambda_{m} m_{AF} \right)\\
- \Lambda \mu
+ 8 \Lambda \lambda_{\chi} \chi_{AB}
+ 8 \Lambda \left( \lambda_{\Delta_S} \Delta_S + \lambda_{\Delta_T} \Delta_T \right),
\label{eq:tildeH_MF_tjU_SGA}
\end{multline}
where $
 \hat\Psi^\dagger_\bk = \begin{pmatrix} \hat c^\dagger_{\bk\uparrow}, & \hat c_{-\bk\downarrow}, & \hat c^\dagger_{\bk+{\bf Q}\uparrow}, & \hat c_{-\bk+{\bf Q}\downarrow} \end{pmatrix}
$, the sum is evaluated over the reduced (magnetic) Brillouin zone (${|k_x| + |k_y| \leqslant \pi}$),
and
\begin{widetext}
\begin{equation}
\hspace{-0.15\textwidth}
  \mathbbm{\tilde M}_\bk = \begin{pmatrix} %
    -\lambda_{\chi} \epsilon_{\bk} - \frac{1}{2} \lambda_n - \mu &
    -\lambda_{\Delta_S} \eta_{\bk}&
    - \frac{1}{2} \lambda_{m} &
    \lambda_{\Delta_T} \eta_{\bk}\\
    -\lambda_{\Delta_S} \eta_{\bk} &
    \lambda_{\chi} \epsilon_{\bk} + \frac{1}{2} \lambda_n + \mu &
    -\lambda_{\Delta_T} \eta_{\bk}&
    - \frac{1}{2} \lambda_{m} \\
    - \frac{1}{2} \lambda_{m} &
    -\lambda_{\Delta_T} \eta_{\bk}&
    \lambda_{\chi} \epsilon_{\bk} - \frac{1}{2} \lambda_n - \mu &
    \lambda_{\Delta_S} \eta_{\bk} \\
    \lambda_{\Delta_T} \eta_{\bk}&
    - \frac{1}{2} \lambda_{m} &
    \lambda_{\Delta_S} \eta_{\bk} &
    -\lambda_{\chi} \epsilon_{\bk} + \frac{1}{2} \lambda_n + \mu
\end{pmatrix}, \label{eq:macierzMM}
\end{equation}
\end{widetext}
where for the square lattice
\begin{subequations}
\begin{eqnarray}
 \epsilon_{\bk} & \equiv & 2 \left( \cos k_x + \cos k_y\right), \\
 \eta_{\bk} & \equiv & 2 \left( \cos k_x - \cos k_y\right).
\end{eqnarray}
\end{subequations}
Diagonalization of $\mathbbm{\tilde M}_\bk$ yields four branches of eigenvalues with their explicit form
\begin{eqnarray}
E_{l\bk} \equiv E_{\pm\, \pm\, \bk} & = & \pm \frac{1}{2} \sqrt{K_{1\bk} \pm 2 \sqrt{K_{2\bk}}},
\end{eqnarray}
where $l=1,\ldots,4$, and
\begin{subequations}
\begin{eqnarray}
K_{1\bk} & \equiv & 4 \epsilon_{\bk}^2 \lambda_{\chi}^2 +  \left(\lambda_{n} + 2 \mu\right)^2 \nonumber\\\hspace{24pt}
&& + 4  \eta_{\bk}^2 \left(\lambda_{\Delta_S}^2 + \lambda_{\Delta_T}^2 \right) + \lambda_{m}^2, \label{eq:K1}\\
K_{2\bk} & \equiv &
\left( 4 \eta_{\bk}^2 \lambda_{\Delta_S} \lambda_{\Delta_T} + \lambda_{m} \left( \lambda_n + 2\mu\right)\right)^2  \nonumber\\\hspace{24pt}
& & +\ 4 \epsilon_{\bk}^2 \lambda_{\chi}^2 \left(4 \eta_{\bk}^2 \lambda_{\Delta_T}^2 + \left(\lambda_{n} + 2 \mu\right)^2 \right).\label{eq:K2}
\end{eqnarray}
\end{subequations}
The energies $\{E_{l\bk}\}_{l=1,\ldots,4}$ represent quasiparticle bands 
after all parameters (mean-fields parameters, the Lagrange multipliers, and $d$) are determined variationally.

The generalized grand potential functional at temperature $T>0$ as given by
 \begin{equation}
 \mathcal{F} = -\frac{1}{\beta} \ln{\mathcal{Z}}, \hspace{6pt} \mbox{with} \hspace{6pt}  \mathcal{Z} = \mathrm{Tr}\left( e^{-\beta \hat K} \right), \label{eq:Funkcjonal_SGA_def2}
\end{equation}
and $\beta \equiv 1/k_B T$, thus
\begin{multline}
 \mathcal{F} / \Lambda= 
8 g_t t \chi_{AB} 
- g_s J \left( \frac{1}{2} m_{AF}^2 + 3 \chi_{AB}^2 + 3 \Delta_S^2 - \Delta_T^2 \right) \\
+ \frac{1}{2} \lambda_n(n-1) + \frac{1}{2} \lambda_{m} m_{AF}
+ 8 \left( \lambda_{\chi} \chi_{AB} + \lambda_{\Delta_S} \Delta_S + \lambda_{\Delta_T} \Delta_T \right)\\
-\frac{1}{\Lambda \beta} \sum_{l,\bk} \ln(1 + e^{-\beta E_{l\bk}})
+ U d^2 - \mu.
\label{eq:Funkcjonal_SGA-pelny2}
\end{multline}


\section{Explicit form of the conditions for the minimum of $\mathcal{F}$}
\label{app:Equations-explicite}

The necessary conditions for the minimum of $\mathcal{F}$, subject to all constraints (introduced in Eq.~\bref{eq:K}) are
\begin{equation}
 \frac{\partial \mathcal{F}}{\partial \vec A}=0, \hspace{6pt}  \hspace{6pt} \frac{\partial \mathcal{F}}{\partial \vec \lambda}=0, \hspace{6pt} \mbox{and} \hspace{6pt} \frac{\partial \mathcal{F}}{\partial d}=0,\label{eq:dFdl_dFdA2}
\end{equation}
where the five mean-field parameters are labeled collectively as $\vec A$, the five Lagrange multipliers as $\vec \lambda$, and $d^2$ is double occupancy probability. In explicit form $\partial \mathcal{F}/\partial \vec A=0$ stands for
\begin{subequations}
\allowdisplaybreaks
\begin{eqnarray}
 && \lambda_{\chi} = -g_t t + \frac{3}{4} g_s J \chi_{AB}, \label{eq:conditions-minimize-first2}\\
 && \lambda_{\Delta_S} = \frac{3}{4} g_s J \Delta_S, \\
 && \lambda_{\Delta_T} = -\frac{1}{4} g_s J \Delta_T, \\
 && \lambda_n = -16 t \chi_{AB} \frac{\partial g_t}{\partial n}
\nonumber\\&& \hspace{12pt}
- 2J \left(-\frac{1}{2} m_{AF}^2 - 3 \chi_{AB}^2 
- 3 \Delta_S^2 + \Delta_T^2 \right) \frac{\partial g_s}{\partial n}, \\
 && \lambda_{m} = 2g_sJ m_{AF} -16 t \chi_{AB} \frac{\partial g_t}{\partial m_{AF}}
\nonumber\\&& \hspace{12pt}
- 2J \left(-\frac{1}{2} m_{AF}^2 - 3 \chi_{AB}^2 
- 3 \Delta_S^2 + \Delta_T^2 \right) \frac{\partial g_s}{\partial m_{AF}},
\label{eq:conditions-minimize-last2}
\end{eqnarray}
\end{subequations}
$\partial \mathcal{F}/\partial \vec \lambda=0$ can be evaluated as
\begin{subequations}
\allowdisplaybreaks
\begin{eqnarray}
 && \frac{1}{\Lambda} \sum_{\bk,l} f(E_{l\bk})\, \partial_{\lambda_{\chi}}{E_{l\bk}}
+ 8 \chi_{AB} = 0, \label{eq:conditions-minimize-first} \\
 && \frac{1}{\Lambda} \sum_{\bk,l} f(E_{l\bk})\, \partial_{\lambda_{\Delta_S}}{E_{l\bk}}
+ 8 \Delta_S = 0, \\  
 && \frac{1}{\Lambda} \sum_{\bk,l} f(E_{l\bk})\, \partial_{\lambda_{\Delta_T}}{E_{l\bk}}
+ 8 \Delta_T = 0, \\ 
 && \frac{1}{\Lambda} \sum_{\bk,l} f(E_{l\bk})\, \partial_{\lambda_{n}}{E_{l\bk}}
- \frac{1}{2} (1 - n) = 0, \\
 && \frac{1}{\Lambda} \sum_{\bk,l} f(E_{l\bk})\, \partial_{\lambda_{m}}{E_{l\bk}}
+ \frac{1}{2} m_{AF} = 0,
\label{eq:conditions-minimize-last}
\end{eqnarray}
\end{subequations}
and $\partial \mathcal{F}/\partial d=0$ denotes
\begin{subequations}
\begin{eqnarray}
&& 2Ud + 8 t \chi_{AB} \frac{\partial g_t}{\partial d} \nonumber\\&& \hspace{12pt}
+J \left( -\frac{1}{2} m_{AF}^2 - 3 \chi_{AB}^2
- 3 \Delta_S^2 + \Delta_T^2 \right) \frac{\partial g_s}{\partial d} = 0,
\label{eq:conditions-minimize-last3}
\end{eqnarray}
\end{subequations}
where $f(E_{l\bk}) \equiv 1/\left(1 + e^{\beta E_{l\bk}} \right)$.
Eqs.\ \bref{eq:conditions-minimize-first2}--\bref{eq:conditions-minimize-last2} can be used to eliminate the parameters $\{ \lambda \}$ from the numerical solution procedure, reducing the number of algebraic equations to six. Consequently, we are left with Eqs.\ \bref{eq:conditions-minimize-first}--\bref{eq:conditions-minimize-last} (the conditions ${\partial \mathcal{F}/\partial\vec\lambda=0}$) and Eq.~\bref{eq:conditions-minimize-last3} ($\partial \mathcal{F}/\partial d=0$).


\begin{table}[t]
\begin{center}
{

\caption{Values of the parameters obtained for the SC phase ($U/|t|=5$ and $\delta = 0.3$) (example 1), for  SC phase ($U/|t|=12$ and $\delta = 0.03$) (example 2), and for the AF+SC phase ($U/|t|=12$ and $\delta=0.001$). The calculations were made for the lattice with $\Lambda=1024\times 1024$ sites. The numerical accuracy is at the last digit specified.}
\label{table:onePoint-pion}
\vspace{12pt}

\begin{tabular}{|c|c|c|c|}
\hline\hline
Variable & SC (1) & SC (2) & AF+SC \\\hline
$\chi_{AB}$			& $0.1907587$			&	$0.1887189$			& $0.1693210$	\\
$\Delta_S$			& $0.00027$			&	$0.138176$			& $0.166906$	\\
$\Delta_T$			& $0$				&	$0$				& $3.92\cdot 10^{-5}$	\\
$\mu$				& $0.5664$			&	$3.5570$			& $3.37154$	\\
$m_{AF}$			& $0$				&	$0$				& $0.13194$	\\
$d^2$				& $5.22266 \cdot 10^{-2}$	&	$8.16196\cdot 10^{-3}$		& $2.2406\cdot 10^{-4}$	\\
$\lambda_{\chi}$		& $0.9661403$			&	$0.327769$			& $0.168074$	\\
$\lambda_{\Delta_S}$		& $0$				&	$0.1258974$			& $0.160777$	\\
$\lambda_{\Delta_T}$		& $0$				&	$0$				& $1.2595\cdot 10^{-5}$	\\
$\lambda_{n}$			& $-2.526087$			&	$-7.176836$			& $-6.744724$	\\
$\lambda_{m}$			& $0$				&	$0$				& $0.100911$	\\
$W$				& $-1.150925$			&	$-0.33669191$			& $-0.233031$	\\
$g_t$				& $0.884438$			&	$0.1558210$			& $4.97139 \cdot 10^{-3}$	\\
$g_s$				& $1.713202$			&	$3.644549$			& $3.85310$	\\
$g_\Delta$			& $0.884438$			&	$0.1558210$			& $4.99912 \cdot 10^{-3}$	\\
$g_m$				& $1.3088937$			&	$1.9090702$			& $1.96293$	\\
$\chi^{c}_{AB}$			& $0.1687143$			&	$2.94064 \cdot 10^{-2}$		& $8.41761 \cdot 10^{-4}$	\\
$\Delta^{c}_S$			& $0.00024$			&	$2.15306 \cdot 10^{-2}$		& $8.343884 \cdot 10^{-4}$	\\
$\Delta^{c}_T$			& $0$				&	$0$				& $1.960 \cdot 10^{-8}$	\\
$m^{c}_{AF}$			& $0$				&	$0$				& $0.25900$ \\\hline\hline
\end{tabular}
}
\end{center}
\end{table}



\section{An alternative procedure of introducing the constraints via Lagrange multipliers}
\label{app:2-schemas-discussion}

In the main text we work with the mean-field grand Hamiltonian $\hat K$, defined as $\hat K \equiv W - \sum_\iota \left( \lambda_\iota \left( \hat O_\iota - O_\iota \right) + \mbox{H.c.}\right) - \mu \hat N$, where $W \equiv \left\langle \mathcal{\hat H}_{\mathit{eff}} \right\rangle_0$ (cf.~Eqs.~\bref{eq:H_eff_tJU} and ~\bref{eq:W}), and $\{ \hat O_\iota \}$ are those operators, whose averages are used to construct $W$. Lagrange multipliers $\lambda_\iota$ are introduced to ensure self-consistency of the solution, i.e., $O_\iota \equiv \left\langle \hat O_\iota \right\rangle_0$ (cf. Eq.~\bref{eq:K}).

Next, in order to find optimal (equilibrium) values of mean fields, the grand potential functional $\mathcal{F} = -{\beta}^{-1} \ln{\mathcal{Z}}$,  where $\mathcal{Z} = \mathrm{Tr}\left( e^{-\beta \hat K} \right)$ (cf. Eq.~\bref{eq:Funkcjonal_SGA_def}) is subsequently minimized with respect to mean-fields subject to constraints included in $\hat K$.

An alternative procedure to the one sketched above is to add the self-consistency preserving constraints directly to $\mathcal{\hat H}_{\mathit{eff}}^{\mathit{MF}}$, i.e., to the mean-field approximated $\mathcal{\hat H}_{\mathit{eff}}$. In this formulation, we have again a separate Lagrange multiplier $\lambda_\iota'$ for each mean-field average $O_\iota' \equiv \left\langle \hat O_\iota' \right\rangle_0$ present in $\mathcal{\hat H}_{\mathit{eff}}^{\mathit{MF}}$.
In effect, we construct the effective mean-field Hamiltonian of the form $\mathcal{\hat H}_\lambda \equiv \mathcal{\hat H}_{\textit{eff}}^{\textit{MF}} - \sum_\iota \left( \lambda_\iota' \left( \hat{O}_\iota' - O_\iota' \right) + \mbox{H.c.}\right)$ and the corresponding mean-field grand Hamiltonian  $\hat K' \equiv \mathcal{\hat H}_\lambda - \mu \hat N$. 
As a next step, the functional $\mathcal{F'}$ is constructed (exactly as discussed above). It should be noted, that minimization of $\mathcal{F'}$ subject to constraints included in $\mathcal{\hat H}_\lambda$, leads to a  set of equations {\it different} than  Eqs.~\bref{eq:conditions-minimize-first2}--\bref{eq:conditions-minimize-last3}. However, {\it those two procedures are equivalent}, i.e., the optimal (equilibrium) values of the mean-fields, corresponding to the minimum of $\mathcal{F}$ and $\mathcal{F'}$ (subject to the same constraints), coincide. A difference in the results may occur only for the values of the Lagrange multipliers, but this does not affect the equilibrium values of the calculated physical quantities. Hence, the two approaches are formally equivalent, which can be shown analytically and has also been verified numerically. 

Those two approaches differ also with respect to numerical execution. Namely, within the first procedure, we can easily find the functional dependence of Lagrange multipliers $\vec\lambda_\iota$ on mean fields $O_\iota$ (as shown in Appendix \ref{app:Equations-explicite}). As a result, the number of equations to be solved numerically is reduced by a factor of $2$. In the second approach discussed here, the corresponding  equations for  $\vec\lambda_\iota'$ are much more complicated and it is not possible to solve them analytically. Therefore, one cannot reduce the effort and numerical cost of solving the model at the same time. So, even though the latter method appears more intuitively appealing, as being more similar to the standard mean-field approach, we have used the former method in the discussion in the main text.


\section{Supplementary informations}

For the sake of completeness [cf. Table~\ref{table:onePoint-pion}] we provide the representative values of the parameters calculated for the following phases: SC for ($U/|t|=5$, $\delta = 0.1$, and $U/|t|=12$, $\delta = 0.03$), and AF+SC ($U/|t|=12$, $\delta=0.001$).
The energies in the columns should not be compared directly, as they correspond to different sets of microscopic parameters.
Numerical accuracy is at the level of the last digit specified.

\bibliographystyle{apsrev}


\bibliography{article-tJU-SGA}

\begin{thebibliography}{65}
\expandafter\ifx\csname natexlab\endcsname\relax\def\natexlab#1{#1}\fi
\expandafter\ifx\csname bibnamefont\endcsname\relax
  \def\bibnamefont#1{#1}\fi
\expandafter\ifx\csname bibfnamefont\endcsname\relax
  \def\bibfnamefont#1{#1}\fi
\expandafter\ifx\csname citenamefont\endcsname\relax
  \def\citenamefont#1{#1}\fi
\expandafter\ifx\csname url\endcsname\relax
  \def\url#1{\texttt{#1}}\fi
\expandafter\ifx\csname urlprefix\endcsname\relax\def\urlprefix{URL }\fi
\providecommand{\bibinfo}[2]{#2}
\providecommand{\eprint}[2][]{\url{#2}}

\bibitem[{\citenamefont{Spałek and
  Oleś}(1977)}]{Spatek1977-PhysicaBC.86-88.375}
\bibinfo{author}{\bibfnamefont{J.}~\bibnamefont{Spałek}} \bibnamefont{and}
  \bibinfo{author}{\bibfnamefont{A.}~\bibnamefont{Oleś}},
  \bibinfo{journal}{Physica B+C} \textbf{\bibinfo{volume}{86–88}},
  \bibinfo{pages}{375 } (\bibinfo{year}{1977}).

\bibitem[{\citenamefont{Chao et~al.}(1977)\citenamefont{Chao, Spałek, and
  Oleś}}]{ChaoSpalekOles1977-JPhysC.10.L271}
\bibinfo{author}{\bibfnamefont{K.~A.} \bibnamefont{Chao}},
  \bibinfo{author}{\bibfnamefont{J.}~\bibnamefont{Spałek}}, \bibnamefont{and}
  \bibinfo{author}{\bibfnamefont{A.~M.} \bibnamefont{Oleś}},
  \bibinfo{journal}{J. Phys. C} \textbf{\bibinfo{volume}{10}},
  \bibinfo{pages}{L271} (\bibinfo{year}{1977}).

\bibitem[{\citenamefont{Spałek}(2007)}]{Spalek2007-ActaPhysPolonA.111.409}
\bibinfo{author}{\bibfnamefont{J.}~\bibnamefont{Spałek}},
  \bibinfo{journal}{Acta Phys. Polon. A} \textbf{\bibinfo{volume}{111}},
  \bibinfo{pages}{409} (\bibinfo{year}{2007}).

\bibitem[{\citenamefont{Anderson}(1988)}]{Anderson1988}
\bibinfo{author}{\bibfnamefont{P.~W.} \bibnamefont{Anderson}}, in
  \emph{\bibinfo{booktitle}{Frontiers and borderlines in many-particle
  physics}}, edited by \bibinfo{editor}{\bibfnamefont{R.~A.}
  \bibnamefont{Broglia}} \bibnamefont{and}
  \bibinfo{editor}{\bibfnamefont{J.~R.} \bibnamefont{Schrieffer}}
  (\bibinfo{publisher}{North-Holland}, \bibinfo{address}{Amsterdam},
  \bibinfo{year}{1988}), pp. \bibinfo{pages}{1--40}.

\bibitem[{\citenamefont{Dagotto}(1994)}]{Dagotto1994-RevModPhys.66.763}
\bibinfo{author}{\bibfnamefont{E.}~\bibnamefont{Dagotto}},
  \bibinfo{journal}{Rev. Mod. Phys.} \textbf{\bibinfo{volume}{66}},
  \bibinfo{pages}{763} (\bibinfo{year}{1994}).

\bibitem[{\citenamefont{Sarker et~al.}(1994)\citenamefont{Sarker, Jayaprakash,
  and Krishnamurthy}}]{Sarker1994-PhysC}
\bibinfo{author}{\bibfnamefont{S.~K.} \bibnamefont{Sarker}},
  \bibinfo{author}{\bibfnamefont{C.}~\bibnamefont{Jayaprakash}},
  \bibnamefont{and} \bibinfo{author}{\bibfnamefont{H.~R.}
  \bibnamefont{Krishnamurthy}}, \bibinfo{journal}{Phys. C. Supercond.}
  \textbf{\bibinfo{volume}{228}}, \bibinfo{pages}{309} (\bibinfo{year}{1994}).

\bibitem[{\citenamefont{Sarker and Lovorn}(2010)}]{Sarker2010-PRB.82.014504}
\bibinfo{author}{\bibfnamefont{S.~K.} \bibnamefont{Sarker}} \bibnamefont{and}
  \bibinfo{author}{\bibfnamefont{T.}~\bibnamefont{Lovorn}},
  \bibinfo{journal}{Phys. Rev. B} \textbf{\bibinfo{volume}{82}},
  \bibinfo{pages}{014504} (\bibinfo{year}{2010}).

\bibitem[{\citenamefont{Johnson et~al.}(2001)\citenamefont{Johnson, Fedorov,
  and Valla}}]{Johnson2001-JournOfElectrSpectros.117.153}
\bibinfo{author}{\bibfnamefont{P.}~\bibnamefont{Johnson}},
  \bibinfo{author}{\bibfnamefont{A.}~\bibnamefont{Fedorov}}, \bibnamefont{and}
  \bibinfo{author}{\bibfnamefont{T.}~\bibnamefont{Valla}}, \bibinfo{journal}{J.
  Electron. Spectrosc. Relat. Phenom.} \textbf{\bibinfo{volume}{117}},
  \bibinfo{pages}{153} (\bibinfo{year}{2001}).

\bibitem[{\citenamefont{Lee et~al.}(2006)\citenamefont{Lee, Nagaosa, and
  Wen}}]{Lee2006-RevModPhys.78.17}
\bibinfo{author}{\bibfnamefont{P.~A.} \bibnamefont{Lee}},
  \bibinfo{author}{\bibfnamefont{N.}~\bibnamefont{Nagaosa}}, \bibnamefont{and}
  \bibinfo{author}{\bibfnamefont{X.-G.} \bibnamefont{Wen}},
  \bibinfo{journal}{Rev. Mod. Phys.} \textbf{\bibinfo{volume}{78}},
  \bibinfo{pages}{17} (\bibinfo{year}{2006}).

\bibitem[{\citenamefont{Imada et~al.}(1998)\citenamefont{Imada, Fujimori, and
  Tokura}}]{Imada1998-RevModPhys.70.1039}
\bibinfo{author}{\bibfnamefont{M.}~\bibnamefont{Imada}},
  \bibinfo{author}{\bibfnamefont{A.}~\bibnamefont{Fujimori}}, \bibnamefont{and}
  \bibinfo{author}{\bibfnamefont{Y.}~\bibnamefont{Tokura}},
  \bibinfo{journal}{Rev. Mod. Phys.} \textbf{\bibinfo{volume}{70}},
  \bibinfo{pages}{1039} (\bibinfo{year}{1998}).

\bibitem[{\citenamefont{Damascelli et~al.}(2003)\citenamefont{Damascelli,
  Hussain, and Shen}}]{Damascelli2003-RevModPhys.75.473}
\bibinfo{author}{\bibfnamefont{A.}~\bibnamefont{Damascelli}},
  \bibinfo{author}{\bibfnamefont{Z.}~\bibnamefont{Hussain}}, \bibnamefont{and}
  \bibinfo{author}{\bibfnamefont{Z.-X.} \bibnamefont{Shen}},
  \bibinfo{journal}{Rev. Mod. Phys.} \textbf{\bibinfo{volume}{75}},
  \bibinfo{pages}{473} (\bibinfo{year}{2003}).

\bibitem[{\citenamefont{Tanaka et~al.}(206)\citenamefont{Tanaka, Lee, Lu,
  Fujimori, Fujii, Risdiana, Terasaki, Scalapino, Devereaux, Hussain
  et~al.}}]{Tanaka2006-Science.314.1910}
\bibinfo{author}{\bibfnamefont{K.}~\bibnamefont{Tanaka}},
  \bibinfo{author}{\bibfnamefont{W.~S.} \bibnamefont{Lee}},
  \bibinfo{author}{\bibfnamefont{D.~H.} \bibnamefont{Lu}},
  \bibinfo{author}{\bibfnamefont{A.}~\bibnamefont{Fujimori}},
  \bibinfo{author}{\bibfnamefont{T.}~\bibnamefont{Fujii}},
  \bibinfo{author}{\bibnamefont{Risdiana}},
  \bibinfo{author}{\bibfnamefont{I.}~\bibnamefont{Terasaki}},
  \bibinfo{author}{\bibfnamefont{D.~J.} \bibnamefont{Scalapino}},
  \bibinfo{author}{\bibfnamefont{T.~P.} \bibnamefont{Devereaux}},
  \bibinfo{author}{\bibfnamefont{Z.}~\bibnamefont{Hussain}},
  \bibnamefont{et~al.}, \bibinfo{journal}{Science}
  \textbf{\bibinfo{volume}{314}}, \bibinfo{pages}{1910} (\bibinfo{year}{206}).

\bibitem[{\citenamefont{Scalapino}(2012)}]{Scalapino2012-RevModPhys.84.1383}
\bibinfo{author}{\bibfnamefont{D.~J.} \bibnamefont{Scalapino}},
  \bibinfo{journal}{Rev. Mod. Phys.} \textbf{\bibinfo{volume}{84}},
  \bibinfo{pages}{1383} (\bibinfo{year}{2012}).

\bibitem[{\citenamefont{Zhang and Rice}(1988)}]{Zhang1988-PhysRevB.37.3759}
\bibinfo{author}{\bibfnamefont{F.~C.} \bibnamefont{Zhang}} \bibnamefont{and}
  \bibinfo{author}{\bibfnamefont{T.~M.} \bibnamefont{Rice}},
  \bibinfo{journal}{Phys. Rev. B} \textbf{\bibinfo{volume}{37}},
  \bibinfo{pages}{3759} (\bibinfo{year}{1988}).

\bibitem[{\citenamefont{Zaanen and Sawatzky}(1990)}]{Zaanen19990-JSSC.88.8}
\bibinfo{author}{\bibfnamefont{J.}~\bibnamefont{Zaanen}} \bibnamefont{and}
  \bibinfo{author}{\bibfnamefont{G.}~\bibnamefont{Sawatzky}},
  \bibinfo{journal}{J. Solid State Chem.} \textbf{\bibinfo{volume}{88}},
  \bibinfo{pages}{8 } (\bibinfo{year}{1990}).

\bibitem[{\citenamefont{Daul et~al.}(2000)\citenamefont{Daul, Scalapino, and
  White}}]{Daul2000-PRL.84.4188}
\bibinfo{author}{\bibfnamefont{S.}~\bibnamefont{Daul}},
  \bibinfo{author}{\bibfnamefont{D.~J.} \bibnamefont{Scalapino}},
  \bibnamefont{and} \bibinfo{author}{\bibfnamefont{S.~R.} \bibnamefont{White}},
  \bibinfo{journal}{Phys. Rev. Lett.} \textbf{\bibinfo{volume}{84}},
  \bibinfo{pages}{4188} (\bibinfo{year}{2000}).

\bibitem[{\citenamefont{Basu et~al.}(2001)\citenamefont{Basu, Gooding, and
  Leung}}]{Basu2001-PRB.63.100506}
\bibinfo{author}{\bibfnamefont{S.}~\bibnamefont{Basu}},
  \bibinfo{author}{\bibfnamefont{R.~J.} \bibnamefont{Gooding}},
  \bibnamefont{and} \bibinfo{author}{\bibfnamefont{P.~W.} \bibnamefont{Leung}},
  \bibinfo{journal}{Phys. Rev. B} \textbf{\bibinfo{volume}{63}},
  \bibinfo{pages}{100506} (\bibinfo{year}{2001}).

\bibitem[{\citenamefont{Zhang}(2003)}]{Zhang2003-prl}
\bibinfo{author}{\bibfnamefont{F.~C.} \bibnamefont{Zhang}},
  \bibinfo{journal}{Phys. Rev. Lett.} \textbf{\bibinfo{volume}{90}},
  \bibinfo{pages}{207002} (\bibinfo{year}{2003}).

\bibitem[{\citenamefont{Xiang et~al.}(2009)\citenamefont{Xiang, Luo, Lu, Shen,
  and Shen}}]{Xiang2009-PRB.79.014524}
\bibinfo{author}{\bibfnamefont{T.}~\bibnamefont{Xiang}},
  \bibinfo{author}{\bibfnamefont{H.~G.} \bibnamefont{Luo}},
  \bibinfo{author}{\bibfnamefont{D.~H.} \bibnamefont{Lu}},
  \bibinfo{author}{\bibfnamefont{K.~M.} \bibnamefont{Shen}}, \bibnamefont{and}
  \bibinfo{author}{\bibfnamefont{Z.~X.} \bibnamefont{Shen}},
  \bibinfo{journal}{Phys. Rev. B} \textbf{\bibinfo{volume}{79}},
  \bibinfo{pages}{014524} (\bibinfo{year}{2009}).

\bibitem[{\citenamefont{Yu et~al.}(2007)\citenamefont{Yu, Higgins, Bach, and
  Greene}}]{Yu2007-PhysRevB.76.020503}
\bibinfo{author}{\bibfnamefont{W.}~\bibnamefont{Yu}},
  \bibinfo{author}{\bibfnamefont{J.~S.} \bibnamefont{Higgins}},
  \bibinfo{author}{\bibfnamefont{P.}~\bibnamefont{Bach}}, \bibnamefont{and}
  \bibinfo{author}{\bibfnamefont{R.~L.} \bibnamefont{Greene}},
  \bibinfo{journal}{Phys. Rev. B} \textbf{\bibinfo{volume}{76}},
  \bibinfo{pages}{020503} (\bibinfo{year}{2007}).

\bibitem[{\citenamefont{Armitage et~al.}(2010)\citenamefont{Armitage, Fournier,
  and Greene}}]{Armitage2010-RevModPhys.82.2421}
\bibinfo{author}{\bibfnamefont{N.~P.} \bibnamefont{Armitage}},
  \bibinfo{author}{\bibfnamefont{P.}~\bibnamefont{Fournier}}, \bibnamefont{and}
  \bibinfo{author}{\bibfnamefont{R.~L.} \bibnamefont{Greene}},
  \bibinfo{journal}{Rev. Mod. Phys.} \textbf{\bibinfo{volume}{82}},
  \bibinfo{pages}{2421} (\bibinfo{year}{2010}).

\bibitem[{\citenamefont{Kimura et~al.}(1999)\citenamefont{Kimura, Hirota,
  Matsushita, Yamada, Endoh, Lee, Majkrzak, Erwin, Shirane, Greven
  et~al.}}]{Kimura1999-PhysRevB.59.6517}
\bibinfo{author}{\bibfnamefont{H.}~\bibnamefont{Kimura}},
  \bibinfo{author}{\bibfnamefont{K.}~\bibnamefont{Hirota}},
  \bibinfo{author}{\bibfnamefont{H.}~\bibnamefont{Matsushita}},
  \bibinfo{author}{\bibfnamefont{K.}~\bibnamefont{Yamada}},
  \bibinfo{author}{\bibfnamefont{Y.}~\bibnamefont{Endoh}},
  \bibinfo{author}{\bibfnamefont{S.-H.} \bibnamefont{Lee}},
  \bibinfo{author}{\bibfnamefont{C.~F.} \bibnamefont{Majkrzak}},
  \bibinfo{author}{\bibfnamefont{R.}~\bibnamefont{Erwin}},
  \bibinfo{author}{\bibfnamefont{G.}~\bibnamefont{Shirane}},
  \bibinfo{author}{\bibfnamefont{M.}~\bibnamefont{Greven}},
  \bibnamefont{et~al.}, \bibinfo{journal}{Phys. Rev. B}
  \textbf{\bibinfo{volume}{59}}, \bibinfo{pages}{6517} (\bibinfo{year}{1999}).

\bibitem[{\citenamefont{Lee et~al.}(1999)\citenamefont{Lee, Birgeneau, Kastner,
  Endoh, Wakimoto, Yamada, Erwin, Lee, and Shirane}}]{Lee1999-PhysRevB.60.3643}
\bibinfo{author}{\bibfnamefont{Y.~S.} \bibnamefont{Lee}},
  \bibinfo{author}{\bibfnamefont{R.~J.} \bibnamefont{Birgeneau}},
  \bibinfo{author}{\bibfnamefont{M.~A.} \bibnamefont{Kastner}},
  \bibinfo{author}{\bibfnamefont{Y.}~\bibnamefont{Endoh}},
  \bibinfo{author}{\bibfnamefont{S.}~\bibnamefont{Wakimoto}},
  \bibinfo{author}{\bibfnamefont{K.}~\bibnamefont{Yamada}},
  \bibinfo{author}{\bibfnamefont{R.~W.} \bibnamefont{Erwin}},
  \bibinfo{author}{\bibfnamefont{S.-H.} \bibnamefont{Lee}}, \bibnamefont{and}
  \bibinfo{author}{\bibfnamefont{G.}~\bibnamefont{Shirane}},
  \bibinfo{journal}{Phys. Rev. B} \textbf{\bibinfo{volume}{60}},
  \bibinfo{pages}{3643} (\bibinfo{year}{1999}).

\bibitem[{\citenamefont{Sidis et~al.}(2001)\citenamefont{Sidis, Ulrich,
  Bourges, Bernhard, Niedermayer, Regnault, Andersen, and
  Keimer}}]{Sidis2001-PhysRevLett.86.4100}
\bibinfo{author}{\bibfnamefont{Y.}~\bibnamefont{Sidis}},
  \bibinfo{author}{\bibfnamefont{C.}~\bibnamefont{Ulrich}},
  \bibinfo{author}{\bibfnamefont{P.}~\bibnamefont{Bourges}},
  \bibinfo{author}{\bibfnamefont{C.}~\bibnamefont{Bernhard}},
  \bibinfo{author}{\bibfnamefont{C.}~\bibnamefont{Niedermayer}},
  \bibinfo{author}{\bibfnamefont{L.~P.} \bibnamefont{Regnault}},
  \bibinfo{author}{\bibfnamefont{N.~H.} \bibnamefont{Andersen}},
  \bibnamefont{and} \bibinfo{author}{\bibfnamefont{B.}~\bibnamefont{Keimer}},
  \bibinfo{journal}{Phys. Rev. Lett.} \textbf{\bibinfo{volume}{86}},
  \bibinfo{pages}{4100} (\bibinfo{year}{2001}).

\bibitem[{\citenamefont{Hodges et~al.}(2002)\citenamefont{Hodges, Sidis,
  Bourges, Mirebeau, Hennion, and Chaud}}]{Hodges2002-PhysRevB.66.020501}
\bibinfo{author}{\bibfnamefont{J.~A.} \bibnamefont{Hodges}},
  \bibinfo{author}{\bibfnamefont{Y.}~\bibnamefont{Sidis}},
  \bibinfo{author}{\bibfnamefont{P.}~\bibnamefont{Bourges}},
  \bibinfo{author}{\bibfnamefont{I.}~\bibnamefont{Mirebeau}},
  \bibinfo{author}{\bibfnamefont{M.}~\bibnamefont{Hennion}}, \bibnamefont{and}
  \bibinfo{author}{\bibfnamefont{X.}~\bibnamefont{Chaud}},
  \bibinfo{journal}{Phys. Rev. B} \textbf{\bibinfo{volume}{66}},
  \bibinfo{pages}{020501} (\bibinfo{year}{2002}).

\bibitem[{\citenamefont{Kawamoto et~al.}(2008)\citenamefont{Kawamoto, Bando,
  Mori, Konoike, Takahide, Terashima, Uji, Takimiya, and
  Otsubo}}]{Kawamoto2008-PhysRevB.77.224506}
\bibinfo{author}{\bibfnamefont{T.}~\bibnamefont{Kawamoto}},
  \bibinfo{author}{\bibfnamefont{Y.}~\bibnamefont{Bando}},
  \bibinfo{author}{\bibfnamefont{T.}~\bibnamefont{Mori}},
  \bibinfo{author}{\bibfnamefont{T.}~\bibnamefont{Konoike}},
  \bibinfo{author}{\bibfnamefont{Y.}~\bibnamefont{Takahide}},
  \bibinfo{author}{\bibfnamefont{T.}~\bibnamefont{Terashima}},
  \bibinfo{author}{\bibfnamefont{S.}~\bibnamefont{Uji}},
  \bibinfo{author}{\bibfnamefont{K.}~\bibnamefont{Takimiya}}, \bibnamefont{and}
  \bibinfo{author}{\bibfnamefont{T.}~\bibnamefont{Otsubo}},
  \bibinfo{journal}{Phys. Rev. B} \textbf{\bibinfo{volume}{77}},
  \bibinfo{pages}{224506} (\bibinfo{year}{2008}).

\bibitem[{\citenamefont{Mathur et~al.}(1998)\citenamefont{Mathur, Grosche,
  Julian, Walker, Freye, Haselwimmer, and
  Lonzarich}}]{Mathur1998-2010-Nature.394.39}
\bibinfo{author}{\bibfnamefont{N.~D.} \bibnamefont{Mathur}},
  \bibinfo{author}{\bibfnamefont{F.~M.} \bibnamefont{Grosche}},
  \bibinfo{author}{\bibfnamefont{S.~R.} \bibnamefont{Julian}},
  \bibinfo{author}{\bibfnamefont{I.~R.} \bibnamefont{Walker}},
  \bibinfo{author}{\bibfnamefont{D.~M.} \bibnamefont{Freye}},
  \bibinfo{author}{\bibfnamefont{R.~K.~W.} \bibnamefont{Haselwimmer}},
  \bibnamefont{and} \bibinfo{author}{\bibfnamefont{G.~G.}
  \bibnamefont{Lonzarich}}, \bibinfo{journal}{Nature}
  \textbf{\bibinfo{volume}{394}}, \bibinfo{pages}{39} (\bibinfo{year}{1998}).

\bibitem[{\citenamefont{Ma et~al.}(2012)\citenamefont{Ma, Ji, Dai, Lu, Eom,
  Kim, Normand, and Yu}}]{Ma2012-PhysRevLett.109.197002}
\bibinfo{author}{\bibfnamefont{L.}~\bibnamefont{Ma}},
  \bibinfo{author}{\bibfnamefont{G.~F.} \bibnamefont{Ji}},
  \bibinfo{author}{\bibfnamefont{J.}~\bibnamefont{Dai}},
  \bibinfo{author}{\bibfnamefont{X.~R.} \bibnamefont{Lu}},
  \bibinfo{author}{\bibfnamefont{M.~J.} \bibnamefont{Eom}},
  \bibinfo{author}{\bibfnamefont{J.~S.} \bibnamefont{Kim}},
  \bibinfo{author}{\bibfnamefont{B.}~\bibnamefont{Normand}}, \bibnamefont{and}
  \bibinfo{author}{\bibfnamefont{W.}~\bibnamefont{Yu}}, \bibinfo{journal}{Phys.
  Rev. Lett.} \textbf{\bibinfo{volume}{109}}, \bibinfo{pages}{197002}
  (\bibinfo{year}{2012}).

\bibitem[{\citenamefont{Li et~al.}(2012)\citenamefont{Li, Zhou, Liu, Sun, Yang,
  Lin, and Zheng}}]{Li2012-PhysRevB.86.180501}
\bibinfo{author}{\bibfnamefont{Z.}~\bibnamefont{Li}},
  \bibinfo{author}{\bibfnamefont{R.}~\bibnamefont{Zhou}},
  \bibinfo{author}{\bibfnamefont{Y.}~\bibnamefont{Liu}},
  \bibinfo{author}{\bibfnamefont{D.~L.} \bibnamefont{Sun}},
  \bibinfo{author}{\bibfnamefont{J.}~\bibnamefont{Yang}},
  \bibinfo{author}{\bibfnamefont{C.~T.} \bibnamefont{Lin}}, \bibnamefont{and}
  \bibinfo{author}{\bibfnamefont{G.-q.} \bibnamefont{Zheng}},
  \bibinfo{journal}{Phys. Rev. B} \textbf{\bibinfo{volume}{86}},
  \bibinfo{pages}{180501} (\bibinfo{year}{2012}).

\bibitem[{\citenamefont{Marsik et~al.}(2010)\citenamefont{Marsik, Kim, Dubroka,
  R\"ossle, Malik, Schulz, Wang, Niedermayer, Drew, Willis
  et~al.}}]{Marsik2010-PhysRevLett.105.057001}
\bibinfo{author}{\bibfnamefont{P.}~\bibnamefont{Marsik}},
  \bibinfo{author}{\bibfnamefont{K.~W.} \bibnamefont{Kim}},
  \bibinfo{author}{\bibfnamefont{A.}~\bibnamefont{Dubroka}},
  \bibinfo{author}{\bibfnamefont{M.}~\bibnamefont{R\"ossle}},
  \bibinfo{author}{\bibfnamefont{V.~K.} \bibnamefont{Malik}},
  \bibinfo{author}{\bibfnamefont{L.}~\bibnamefont{Schulz}},
  \bibinfo{author}{\bibfnamefont{C.~N.} \bibnamefont{Wang}},
  \bibinfo{author}{\bibfnamefont{C.}~\bibnamefont{Niedermayer}},
  \bibinfo{author}{\bibfnamefont{A.~J.} \bibnamefont{Drew}},
  \bibinfo{author}{\bibfnamefont{M.}~\bibnamefont{Willis}},
  \bibnamefont{et~al.}, \bibinfo{journal}{Phys. Rev. Lett.}
  \textbf{\bibinfo{volume}{105}}, \bibinfo{pages}{057001}
  (\bibinfo{year}{2010}).

\bibitem[{\citenamefont{Bernhard et~al.}(2012)\citenamefont{Bernhard, Wang,
  Nuccio, Schulz, Zaharko, Larsen, Aristizabal, Willis, Drew, Varma
  et~al.}}]{Bernhard2012-PhysRevB.86.184509}
\bibinfo{author}{\bibfnamefont{C.}~\bibnamefont{Bernhard}},
  \bibinfo{author}{\bibfnamefont{C.~N.} \bibnamefont{Wang}},
  \bibinfo{author}{\bibfnamefont{L.}~\bibnamefont{Nuccio}},
  \bibinfo{author}{\bibfnamefont{L.}~\bibnamefont{Schulz}},
  \bibinfo{author}{\bibfnamefont{O.}~\bibnamefont{Zaharko}},
  \bibinfo{author}{\bibfnamefont{J.}~\bibnamefont{Larsen}},
  \bibinfo{author}{\bibfnamefont{C.}~\bibnamefont{Aristizabal}},
  \bibinfo{author}{\bibfnamefont{M.}~\bibnamefont{Willis}},
  \bibinfo{author}{\bibfnamefont{A.~J.} \bibnamefont{Drew}},
  \bibinfo{author}{\bibfnamefont{G.~D.} \bibnamefont{Varma}},
  \bibnamefont{et~al.}, \bibinfo{journal}{Phys. Rev. B}
  \textbf{\bibinfo{volume}{86}}, \bibinfo{pages}{184509}
  (\bibinfo{year}{2012}).

\bibitem[{\citenamefont{Milovanovi\ifmmode~\acute{c}\else \'{c}\fi{} and
  Predin}(2012)}]{Milovanovic2012-PhysRevB.86.195113}
\bibinfo{author}{\bibfnamefont{M.~V.}
  \bibnamefont{Milovanovi\ifmmode~\acute{c}\else \'{c}\fi{}}} \bibnamefont{and}
  \bibinfo{author}{\bibfnamefont{S.}~\bibnamefont{Predin}},
  \bibinfo{journal}{Phys. Rev. B} \textbf{\bibinfo{volume}{86}},
  \bibinfo{pages}{195113} (\bibinfo{year}{2012}).

\bibitem[{\citenamefont{Gan et~al.}(2005{\natexlab{a}})\citenamefont{Gan,
  Zhang, and Su}}]{Gan2005-prb}
\bibinfo{author}{\bibfnamefont{J.~Y.} \bibnamefont{Gan}},
  \bibinfo{author}{\bibfnamefont{F.~C.} \bibnamefont{Zhang}}, \bibnamefont{and}
  \bibinfo{author}{\bibfnamefont{Z.~B.} \bibnamefont{Su}},
  \bibinfo{journal}{Phys. Rev. B} \textbf{\bibinfo{volume}{71}},
  \bibinfo{pages}{014508} (\bibinfo{year}{2005}{\natexlab{a}}).

\bibitem[{\citenamefont{Gan et~al.}(2005{\natexlab{b}})\citenamefont{Gan, Chen,
  Su, and Zhang}}]{Gan2005-prl}
\bibinfo{author}{\bibfnamefont{J.~Y.} \bibnamefont{Gan}},
  \bibinfo{author}{\bibfnamefont{Y.}~\bibnamefont{Chen}},
  \bibinfo{author}{\bibfnamefont{Z.~B.} \bibnamefont{Su}}, \bibnamefont{and}
  \bibinfo{author}{\bibfnamefont{F.~C.} \bibnamefont{Zhang}},
  \bibinfo{journal}{Phys. Rev. Lett.} \textbf{\bibinfo{volume}{94}},
  \bibinfo{pages}{067005} (\bibinfo{year}{2005}{\natexlab{b}}).

\bibitem[{\citenamefont{Bernevig et~al.}(2003)\citenamefont{Bernevig, Laughlin,
  and Santiago}}]{Bernevig-PhysRevLett.91.147003}
\bibinfo{author}{\bibfnamefont{B.~A.} \bibnamefont{Bernevig}},
  \bibinfo{author}{\bibfnamefont{R.~B.} \bibnamefont{Laughlin}},
  \bibnamefont{and} \bibinfo{author}{\bibfnamefont{D.~I.}
  \bibnamefont{Santiago}}, \bibinfo{journal}{Phys. Rev. Lett.}
  \textbf{\bibinfo{volume}{91}}, \bibinfo{pages}{147003}
  (\bibinfo{year}{2003}).

\bibitem[{\citenamefont{Laughlin}(2006)}]{Laughlin2006}
\bibinfo{author}{\bibfnamefont{R.~B.} \bibnamefont{Laughlin}},
  \bibinfo{journal}{Philosophical Magazine} \textbf{\bibinfo{volume}{86}},
  \bibinfo{pages}{1165} (\bibinfo{year}{2006}).

\bibitem[{\citenamefont{Van~Harlingen}(1995)}]{Harlingen1995-RevModPhys.67.515}
\bibinfo{author}{\bibfnamefont{D.~J.} \bibnamefont{Van~Harlingen}},
  \bibinfo{journal}{Rev. Mod. Phys.} \textbf{\bibinfo{volume}{67}},
  \bibinfo{pages}{515} (\bibinfo{year}{1995}).

\bibitem[{\citenamefont{Tsuei and Kirtley}(2000)}]{Tsuei2000-RevModPhys.72.969}
\bibinfo{author}{\bibfnamefont{C.~C.} \bibnamefont{Tsuei}} \bibnamefont{and}
  \bibinfo{author}{\bibfnamefont{J.~R.} \bibnamefont{Kirtley}},
  \bibinfo{journal}{Rev. Mod. Phys.} \textbf{\bibinfo{volume}{72}},
  \bibinfo{pages}{969} (\bibinfo{year}{2000}).

\bibitem[{\citenamefont{Yuan et~al.}(2005)\citenamefont{Yuan, Yuan, and
  Ting}}]{Yuan2005-prb}
\bibinfo{author}{\bibfnamefont{F.}~\bibnamefont{Yuan}},
  \bibinfo{author}{\bibfnamefont{Q.}~\bibnamefont{Yuan}}, \bibnamefont{and}
  \bibinfo{author}{\bibfnamefont{C.~S.} \bibnamefont{Ting}},
  \bibinfo{journal}{Phys. Rev. B} \textbf{\bibinfo{volume}{71}},
  \bibinfo{pages}{104505} (\bibinfo{year}{2005}).

\bibitem[{\citenamefont{Heiselberg}(2009)}]{Heiselberg2009-pra}
\bibinfo{author}{\bibfnamefont{H.}~\bibnamefont{Heiselberg}},
  \bibinfo{journal}{Phys. Rev. A} \textbf{\bibinfo{volume}{79}},
  \bibinfo{pages}{063611} (\bibinfo{year}{2009}).

\bibitem[{\citenamefont{Voo}(2011)}]{Voo-2011-JPhysCondensMatter.23.495602}
\bibinfo{author}{\bibfnamefont{K.-K.} \bibnamefont{Voo}}, \bibinfo{journal}{J.
  Phys.: Condens. Matter} \textbf{\bibinfo{volume}{23}},
  \bibinfo{pages}{495602} (\bibinfo{year}{2011}).

\bibitem[{\citenamefont{Liu et~al.}(2012)\citenamefont{Liu, Zhang, Yuan, and
  Xia}}]{Fen2012-CommTheorPhys}
\bibinfo{author}{\bibfnamefont{F.-F.} \bibnamefont{Liu}},
  \bibinfo{author}{\bibfnamefont{Y.}~\bibnamefont{Zhang}},
  \bibinfo{author}{\bibfnamefont{F.}~\bibnamefont{Yuan}}, \bibnamefont{and}
  \bibinfo{author}{\bibfnamefont{L.-H.} \bibnamefont{Xia}},
  \bibinfo{journal}{Communications in Theoretical Physics}
  \textbf{\bibinfo{volume}{57}}, \bibinfo{pages}{727} (\bibinfo{year}{2012}).

\bibitem[{\citenamefont{J\k{e}drak et~al.}(2010)\citenamefont{J\k{e}drak,
  Kaczmarczyk, and Spa\l{}ek}}]{Jedrak2010-arXiv}
\bibinfo{author}{\bibfnamefont{J.}~\bibnamefont{J\k{e}drak}},
  \bibinfo{author}{\bibfnamefont{J.}~\bibnamefont{Kaczmarczyk}},
  \bibnamefont{and}
  \bibinfo{author}{\bibfnamefont{J.}~\bibnamefont{Spa\l{}ek}},
  \bibinfo{journal}{arXiv:cond-mat/1008.0021}  (\bibinfo{year}{2010}),
  \bibinfo{note}{unpublished}.

\bibitem[{\citenamefont{Jędrak and Spa\l{}ek}(2010)}]{Jedrak2010-prb}
\bibinfo{author}{\bibfnamefont{J.}~\bibnamefont{Jędrak}} \bibnamefont{and}
  \bibinfo{author}{\bibfnamefont{J.}~\bibnamefont{Spa\l{}ek}},
  \bibinfo{journal}{Phys. Rev. B} \textbf{\bibinfo{volume}{81}},
  \bibinfo{pages}{073108} (\bibinfo{year}{2010}).

\bibitem[{\citenamefont{Jędrak}(2011)}]{Jedrak2011-PhD}
\bibinfo{author}{\bibfnamefont{J.}~\bibnamefont{Jędrak}}, Ph.D. thesis,
  \bibinfo{school}{Jagiellonian University, Kraków} (\bibinfo{year}{2011}),
  \urlprefix\url{http://th-www.if.uj.edu.pl/ztms/download/phdTheses/Jakub_Jedr%
ak_doktorat.pdf}.

\bibitem[{\citenamefont{Gutzwiller}(1963)}]{Gutzwiller1963-prl}
\bibinfo{author}{\bibfnamefont{M.~C.} \bibnamefont{Gutzwiller}},
  \bibinfo{journal}{Phys. Rev. Lett.} \textbf{\bibinfo{volume}{10}},
  \bibinfo{pages}{159} (\bibinfo{year}{1963}).

\bibitem[{\citenamefont{Gutzwiller}(1965)}]{Gutzwiller1965-pr}
\bibinfo{author}{\bibfnamefont{M.~C.} \bibnamefont{Gutzwiller}},
  \bibinfo{journal}{Phys. Rev.} \textbf{\bibinfo{volume}{137}},
  \bibinfo{pages}{A1726} (\bibinfo{year}{1965}).

\bibitem[{\citenamefont{Ogawa et~al.}(1975)\citenamefont{Ogawa, Kanda, and
  Matsubara}}]{Ogawa1975-ProgTheorPhys.53.614}
\bibinfo{author}{\bibfnamefont{T.}~\bibnamefont{Ogawa}},
  \bibinfo{author}{\bibfnamefont{K.}~\bibnamefont{Kanda}}, \bibnamefont{and}
  \bibinfo{author}{\bibfnamefont{T.}~\bibnamefont{Matsubara}},
  \bibinfo{journal}{Prog. Theor. Phys.} \textbf{\bibinfo{volume}{53}},
  \bibinfo{pages}{614} (\bibinfo{year}{1975}).

\bibitem[{\citenamefont{Zhang et~al.}(1988)\citenamefont{Zhang, Gros, Rice, and
  Shiba}}]{Zhang1988-SuperSciTech.1.36}
\bibinfo{author}{\bibfnamefont{F.~C.} \bibnamefont{Zhang}},
  \bibinfo{author}{\bibfnamefont{C.}~\bibnamefont{Gros}},
  \bibinfo{author}{\bibfnamefont{T.~M.} \bibnamefont{Rice}}, \bibnamefont{and}
  \bibinfo{author}{\bibfnamefont{H.}~\bibnamefont{Shiba}},
  \bibinfo{journal}{Supercond. Sci. Technol.} \textbf{\bibinfo{volume}{1}},
  \bibinfo{pages}{36} (\bibinfo{year}{1988}).

\bibitem[{\citenamefont{Himeda and Ogata}(1999)}]{Himeda1999-PRB.60.R9935}
\bibinfo{author}{\bibfnamefont{A.}~\bibnamefont{Himeda}} \bibnamefont{and}
  \bibinfo{author}{\bibfnamefont{M.}~\bibnamefont{Ogata}},
  \bibinfo{journal}{Phys. Rev. B} \textbf{\bibinfo{volume}{60}},
  \bibinfo{pages}{R9935} (\bibinfo{year}{1999}).

\bibitem[{\citenamefont{Tsonis et~al.}(2008)\citenamefont{Tsonis, Kotetes,
  Varelogiannis, and Littlewood}}]{Tsonis2008-jpcm}
\bibinfo{author}{\bibfnamefont{S.}~\bibnamefont{Tsonis}},
  \bibinfo{author}{\bibfnamefont{P.}~\bibnamefont{Kotetes}},
  \bibinfo{author}{\bibfnamefont{G.}~\bibnamefont{Varelogiannis}},
  \bibnamefont{and} \bibinfo{author}{\bibfnamefont{P.~B.}
  \bibnamefont{Littlewood}}, \bibinfo{journal}{J. Phys.: Condens. Matter}
  \textbf{\bibinfo{volume}{20}}, \bibinfo{pages}{434234}
  (\bibinfo{year}{2008}).

\bibitem[{\citenamefont{Aperis et~al.}(2010)\citenamefont{Aperis,
  Varelogiannis, and Littlewood}}]{Aperis2010-prl}
\bibinfo{author}{\bibfnamefont{A.}~\bibnamefont{Aperis}},
  \bibinfo{author}{\bibfnamefont{G.}~\bibnamefont{Varelogiannis}},
  \bibnamefont{and} \bibinfo{author}{\bibfnamefont{P.~B.}
  \bibnamefont{Littlewood}}, \bibinfo{journal}{Phys. Rev. Lett.}
  \textbf{\bibinfo{volume}{104}}, \bibinfo{pages}{216403}
  (\bibinfo{year}{2010}).

\bibitem[{\citenamefont{Bogoliubov}(1958{\natexlab{a}})}]{Bogoliubov1958}
\bibinfo{author}{\bibfnamefont{N.~N.} \bibnamefont{Bogoliubov}},
  \bibinfo{journal}{Soviet Phys. JETP} \textbf{\bibinfo{volume}{34}},
  \bibinfo{pages}{41} (\bibinfo{year}{1958}{\natexlab{a}}).

\bibitem[{\citenamefont{Bogoliubov}(1958{\natexlab{b}})}]{Bogoliubov1958-USSR}
\bibinfo{author}{\bibfnamefont{N.~N.} \bibnamefont{Bogoliubov}},
  \bibinfo{journal}{Zh. Exp. Teor. Fiz.} \textbf{\bibinfo{volume}{34}},
  \bibinfo{pages}{58} (\bibinfo{year}{1958}{\natexlab{b}}).

\bibitem[{\citenamefont{Bogolyubov et~al.}(1958)\citenamefont{Bogolyubov,
  Tolmachev, and Shirkov}}]{BogoliubovTolmachev1958}
\bibinfo{author}{\bibfnamefont{N.~N.} \bibnamefont{Bogolyubov}},
  \bibinfo{author}{\bibfnamefont{V.~V.} \bibnamefont{Tolmachev}},
  \bibnamefont{and} \bibinfo{author}{\bibfnamefont{D.~V.}
  \bibnamefont{Shirkov}}, \bibinfo{journal}{Fortsch.Phys.}
  \textbf{\bibinfo{volume}{6}}, \bibinfo{pages}{605} (\bibinfo{year}{1958}).

\bibitem[{\citenamefont{Slater}(1951)}]{Slater1951-PhysRev.82.538}
\bibinfo{author}{\bibfnamefont{J.~C.} \bibnamefont{Slater}},
  \bibinfo{journal}{Phys. Rev.} \textbf{\bibinfo{volume}{82}},
  \bibinfo{pages}{538} (\bibinfo{year}{1951}).

\bibitem[{\citenamefont{Kaczmarczyk and Spa\l{}ek}(2011)}]{Kaczmarczyk2011-prb}
\bibinfo{author}{\bibfnamefont{J.}~\bibnamefont{Kaczmarczyk}} \bibnamefont{and}
  \bibinfo{author}{\bibfnamefont{J.}~\bibnamefont{Spa\l{}ek}},
  \bibinfo{journal}{Phys. Rev. B} \textbf{\bibinfo{volume}{84}},
  \bibinfo{pages}{125140} (\bibinfo{year}{2011}).

\bibitem[{\citenamefont{Kaczmarczyk}(2011)}]{Kaczmarczyk2011-PhD}
\bibinfo{author}{\bibfnamefont{J.}~\bibnamefont{Kaczmarczyk}}, Ph.D. thesis,
  \bibinfo{school}{Jagiellonian University, Kraków} (\bibinfo{year}{2011}),
  \urlprefix\url{http://th-www.if.uj.edu.pl/ztms/download/phdTheses/Jan_Kaczma%
rczyk_doktorat.pdf}.

\bibitem[{Note1()}]{Note1}
Note1, \bibinfo{note}{with increase of hole-doping the approximation of zero
  temperature described in the main text became weaker. For bigger $\delta $
  (e.q. for ${U/|t|=12}$ bigger than $0.3$) it starts be insufficient. Taking
  bigger $\beta $ moves the limiting value of $\delta $ just a little. Thus,
  for limit of strong hole-doping other computational techniques have to be
  used (since the correlations are weak in such limit, it can be a basic RMFT
  methods as described in Ref.~\protect \onlinecite {Yuan2005-prb}).}

\bibitem[{\citenamefont{Galassi et~al.}(2009)\citenamefont{Galassi, Davies,
  Theiler, Gough, Jungman, Booth, and Rossi}}]{GSL-manual}
\bibinfo{author}{\bibfnamefont{M.}~\bibnamefont{Galassi}},
  \bibinfo{author}{\bibfnamefont{J.}~\bibnamefont{Davies}},
  \bibinfo{author}{\bibfnamefont{J.}~\bibnamefont{Theiler}},
  \bibinfo{author}{\bibfnamefont{B.}~\bibnamefont{Gough}},
  \bibinfo{author}{\bibfnamefont{P.}~\bibnamefont{Jungman},
  \bibfnamefont{G.~abd~Alken}},
  \bibinfo{author}{\bibfnamefont{M.}~\bibnamefont{Booth}}, \bibnamefont{and}
  \bibinfo{author}{\bibfnamefont{F.}~\bibnamefont{Rossi}},
  \bibinfo{journal}{GNU Scientific Library Reference Manual}
  (\bibinfo{year}{2009}), \bibinfo{note}{3rd ed., Network Theory, Ltd.,
  London}.

\bibitem[{\citenamefont{Bünemann}(2005)}]{Bunemann-jp}
\bibinfo{author}{\bibfnamefont{J.}~\bibnamefont{Bünemann}},
  \bibinfo{journal}{J. Phys.: Condens. Matter} \textbf{\bibinfo{volume}{17}},
  \bibinfo{pages}{3807} (\bibinfo{year}{2005}).

\bibitem[{\citenamefont{Yang}(2009)}]{Kai-Yu-njp}
\bibinfo{author}{\bibfnamefont{K.-Y.} \bibnamefont{Yang}},
  \bibinfo{journal}{New J. Phys.} \textbf{\bibinfo{volume}{11}},
  \bibinfo{pages}{055053} (\bibinfo{year}{2009}).

\bibitem[{\citenamefont{Feng et~al.}(2012)\citenamefont{Feng, Zhao, and
  Huang}}]{Feng2012-PRB.85.054509}
\bibinfo{author}{\bibfnamefont{S.}~\bibnamefont{Feng}},
  \bibinfo{author}{\bibfnamefont{H.}~\bibnamefont{Zhao}}, \bibnamefont{and}
  \bibinfo{author}{\bibfnamefont{Z.}~\bibnamefont{Huang}},
  \bibinfo{journal}{Phys. Rev. B} \textbf{\bibinfo{volume}{85}},
  \bibinfo{pages}{054509} (\bibinfo{year}{2012}).

\bibitem[{\citenamefont{Kubo and
  Uchinami}(1975)}]{Kubo1975-ProgTheorPhys.54.1289}
\bibinfo{author}{\bibfnamefont{K.}~\bibnamefont{Kubo}} \bibnamefont{and}
  \bibinfo{author}{\bibfnamefont{M.}~\bibnamefont{Uchinami}},
  \bibinfo{journal}{Prog. Theor. Phys} \textbf{\bibinfo{volume}{54}},
  \bibinfo{pages}{1289} (\bibinfo{year}{1975}).

\bibitem[{\citenamefont{Korbel et~al.}(2003)\citenamefont{Korbel, Wójcik,
  Klejnberg, Spałek, Acquarone, and Lavagna}}]{Korbel2003}
\bibinfo{author}{\bibfnamefont{P.}~\bibnamefont{Korbel}},
  \bibinfo{author}{\bibfnamefont{W.}~\bibnamefont{Wójcik}},
  \bibinfo{author}{\bibfnamefont{A.}~\bibnamefont{Klejnberg}},
  \bibinfo{author}{\bibfnamefont{J.}~\bibnamefont{Spałek}},
  \bibinfo{author}{\bibfnamefont{M.}~\bibnamefont{Acquarone}},
  \bibnamefont{and} \bibinfo{author}{\bibfnamefont{M.}~\bibnamefont{Lavagna}},
  \bibinfo{journal}{Eur. Phys. J. B} \textbf{\bibinfo{volume}{32}},
  \bibinfo{pages}{315} (\bibinfo{year}{2003}).

\end{thebibliography}

\end{document}